\documentclass[aps,prc,twocolumn,amsmath,amssymb,superscriptaddress,nofootinbib,floatfix]{revtex4-2}

\usepackage{graphicx}
\usepackage{amsmath}
\usepackage{amssymb}
\usepackage{bm}
\usepackage{booktabs}
\usepackage{xcolor}
\usepackage[colorlinks=true, breaklinks=true, linkcolor=blue, citecolor=violet, urlcolor=cyan]{hyperref}
\usepackage{placeins}

\hypersetup{pdftitle={Representation dependence of the neutron-proton kinetic energy in adiabatic deuteron stripping},
            pdfauthor={Jin Lei}}

\begin{document}

\title{Representation dependence of the neutron-proton kinetic energy in adiabatic deuteron stripping}

\author{Jin Lei}
\email[]{jinl@tongji.edu.cn}
\affiliation{School of Physics Science and Engineering, Tongji University, Shanghai 200092, China}
\affiliation{Southern Center for Nuclear-Science Theory (SCNT), Institute of Modern Physics,
Chinese Academy of Sciences, Huizhou 516000, Guangdong Province, China}

\date{\today}

\begin{abstract}
Deuteron stripping reactions such as $(d,p)$ are a main tool for studying the
single-particle structure of nuclei, and they are usually analyzed with the adiabatic
distorted-wave approximation (ADWA). With nonlocal nucleon optical potentials, ADWA cross
sections change by tens of percent, up to a factor of two, depending on which realistic
neutron-proton interaction describes the deuteron. Continuum-discretized calculations change
much less, and exact Faddeev calculations reduce the spread by about a factor of ten. Should
this sensitivity be read as a probe of neutron-proton physics at short distances? One
identified route for it in ADWA is a single number, the neutron-proton kinetic energy averaged
over the region where the two nucleons interact. I change the AV18 interaction only at short
distances, keeping neutron-proton scattering, the deuteron binding energy, and the long-range
tail of the deuteron unchanged. This number then rises from 218 to 314 MeV, more than half of
its spread among five realistic interactions, while the coefficient that sets the low-momentum
shape of the form factor moves by only one percent of its realistic range. The number therefore reflects how
the interaction is written, not anything measured. It enters ADWA through two features that the
other calculations treat differently: ADWA weights the deuteron where the two nucleons are
close, and it assigns the deuteron energy to the pair however fast the two nucleons move. For
this short-range change, an average that keeps the full Gaussian shape of the nonlocal
potential, instead of expanding it to lowest order, rises by only 6 MeV against 96 MeV.
\end{abstract}

\maketitle

\section{Introduction}
\label{sec:intro}

The adiabatic distorted-wave approximation (ADWA) of Johnson and Soper
\cite{JohnsonSoper1970} and Johnson and Tandy \cite{JohnsonTandy1974} is the standard
reaction model for $(d,p)$ and $(p,d)$ analyses. It includes deuteron breakup through the
first Weinberg component of the three-body scattering wave function \cite{Laid1993,Pang2013}, and it
is the model used to extract neutron spectroscopic factors from transfer data
\cite{Tsang2005,Lee2006}; the theory and its applications are reviewed in
Ref.~\cite{TimofeyukJohnson2020}. The same framework has been extended to nonlocal and
energy-dependent nucleon optical potentials
\cite{TimofeyukJohnson2013PRL,TimofeyukJohnson2013,JohnsonTimofeyuk2014}, to three-body
forces induced by target excitation \cite{Dinmore2019,Dinmore2021}, and to chiral
three-nucleon forces acting in the deuteron channel \cite{Timofeyuk2020}. In the adiabatic matrix elements of these calculations the
neutron-proton interaction $V_{np}$ enters through the weight defining the first Weinberg
projection, $\bar\phi_1\propto V_{np}\phi_d$, where $\phi_d$ is the deuteron wave function;
this weight emphasizes neutron-proton separations within the range of $V_{np}$. Spectroscopic and astrophysical analyses that use ADWA inherit
this dependence on $V_{np}$.

Such a dependence was reported for ADWA calculations with nonlocal nucleon potentials
\cite{Bailey2016,Bailey2017erratum,Bailey2017}. For $^{26}$Al$(d,p)^{27}$Al at $E_d=12$ MeV,
the ADWA cross sections for transfer to the $^{27}$Al ground state at zero degrees, computed with
different realistic neutron-proton interactions, spread by about $\pm50$ percent about their
central value \cite{Bailey2016}, as quoted in Ref.~\cite{Deltuva2018}.
Deltuva repeated the calculation with the same nucleon-target potentials but solved the
three-body problem exactly with Faddeev-type equations, and found a spread of about $\pm5$
percent \cite{Deltuva2018}. For the $^{26m}$Al isomer, G\'omez-Ramos and Timofeyuk compared ADWA with a
continuum-discretized calculation that uses the same nucleon-target nonlocality, for Hulth\'en
and Reid soft-core deuterons \cite{GomezRamosTimofeyuk2018}. The ratio of the Reid soft-core to
the Hulth\'en cross section at its maximum, over four final states, is 1.03 to 1.24 in ADWA and
1.00 to 1.04 in the continuum-discretized calculation at $E_d=9.2$ MeV, and 1.69 to 2.24 against
1.05 to 1.14 at 50 MeV.

The mechanism identified in Refs.~\cite{Bailey2016,Bailey2017} is an energy shift. ADWA builds the deuteron-target
potential from the neutron-target and proton-target potentials, which for local potentials are
taken at half the deuteron energy, $E_d/2$. Nonlocal nucleon-target potentials depend on the
nucleon momentum, and the transfer samples the deuteron where $V_{np}$ acts, at small
neutron-proton separations where the relative momentum is large. The equivalent local
potentials must then be taken at a higher energy, $E_d/2+\Delta E$. To leading order in the
nonlocality range $\Delta E=\langle T_{np}\rangle_V/2$ \cite{TimofeyukJohnson2013PRL}, where
\begin{equation}
\langle T_{np}\rangle_V=\frac{\langle\phi_d|V_{np}T|\phi_d\rangle}{\langle\phi_d|V_{np}|\phi_d\rangle}
\label{eq:tnpdef}
\end{equation}
is the neutron-proton relative kinetic energy $T$ averaged with the weight $V_{np}\phi_d$.
Because $V_{np}$ is short ranged, this average emphasizes small separations. For AV18
\cite{Wiringa1995}, $\langle T_{np}\rangle_V=218$ MeV, whereas the ordinary expectation value
$\langle T\rangle=\langle\phi_d|T|\phi_d\rangle$ is 20 MeV. Among the interactions considered
in Refs.~\cite{Bailey2016,Bailey2017}, $\langle T_{np}\rangle_V$ differs by up to a factor of
about three. The leading-order formula overestimates the shift: for the Hulth\'en deuteron,
keeping the full dependence on the nonlocality range lowers it from 57 to 43 MeV
\cite{TimofeyukJohnson2013PRL}. The shifts tabulated in Ref.~\cite{Bailey2017}, obtained with a
lowest-order local-equivalent method, also differ strongly among its four deuteron models. A sensitivity of this kind would make
$(d,p)$ reactions a probe of short-range neutron-proton structure, and it has been proposed
that they could help constrain nucleon-nucleon and three-nucleon forces \cite{Timofeyuk2020}.

The low-momentum content of the neutron-proton factor has long been understood. Buttle
and Goldfarb wrote the transfer amplitude in terms of the finite-range form factor
$D(\mathbf r)=V_{np}\phi_d(\mathbf r)$ and expanded it in the momentum transferred to the
neutron-proton pair \cite{ButtleGoldfarb1964}. The leading term is the zero-range constant
$D_0$, and the first correction is set by the low-momentum slope of the normalized form
factor. Across nine realistic and model interactions, $D_0$ agrees to about one percent and the
inverse square root of the slope coefficient to about four percent
\cite{Bailey2016}. The
interactions differ little in the quantities that control low-momentum transfer and much more
in $\langle T_{np}\rangle_V$, which weights high momenta. The question is whether
that larger difference is information about the neutron-proton interaction.

This paper does not explain why ADWA and the exact calculation differ. That would require
computing how the nonlocal adiabatic potential, the distorted waves, and the cross section
change when $V_{np}$ is changed, and none of these is computed here. It answers a narrower
question that decides how the sensitivity can be read: is $\langle T_{np}\rangle_V$ fixed by
two-body data? If it were, the factor-of-three spread among realistic interactions would trace
back to differences in the two-body properties they reproduce, and $(d,p)$ data analyzed in ADWA
could help discriminate among them, as proposed in Ref.~\cite{Timofeyuk2020}. If it is not, a variation of comparable size can
arise from the representation of a single interaction and is not by itself a measure of
short-range neutron-proton structure. I show that it is not, identify the two features of the
adiabatic projection through which this integral enters ADWA, and contrast them with
continuum-discretized and Faddeev treatments.

To show this, I construct a unitary transformation of the neutron-proton Hamiltonian with the
AV18 interaction in the deuteron channel. It acts only at short distances and preserves the two-body scattering
matrix, the deuteron binding energy, and both asymptotic normalizations, so the transformed interaction is AV18 in another
representation, not a new fit. The general principle, that such integrals are not observables
and change under unitary transformations at fixed observables
\cite{PolyzouGlockle1990,Duguet2015,More2017}, is known; what this paper adds is its
quantitative realization in ADWA.

Section~\ref{sec:identity} defines the form factor and writes $\langle T_{np}\rangle_V$ in
terms of the mean and the variance of the neutron-proton kinetic energy.
Section~\ref{sec:representation} constructs the transformation and compares the changes it
produces in $\langle T_{np}\rangle_V$ and in the finite-range coefficient, which sets the
low-momentum shape of the form factor. Section~\ref{sec:whyadwa} shows how the integral enters the adiabatic
projection and contrasts it with continuum-discretized and Faddeev treatments. Section~\ref{sec:parameter} relates the low-momentum form factor of
realistic coupled-channel deuterons to the asymptotic normalization coefficient (ANC), and
Sec.~\ref{sec:where} examines which momenta standard reaction treatments sample.
Section~\ref{sec:recovery} transforms the nucleon-target interactions consistently, which
induces a three-body term and restores an $S$-projected folded kernel, and
Sec.~\ref{sec:scope} discusses high-momentum probes and the limits of the analysis.

\section{The form factor and the adiabatic structure integral}
\label{sec:identity}

\begin{table}[t]
\caption{Deuteron properties of the five interactions used here, from a direct solution of
Eq.~(\ref{eq:momspace}): binding energy $B_d$, $D$-state probability $P_D$, and the quantities below. $f_D^{(V)}$ is defined by Eq.~(\ref{eq:pdbar}) and
$\langle T_{np}\rangle_V$ is defined by Eq.~(\ref{eq:tnpdef}); $c$ is the finite-range coefficient of
Eq.~(\ref{eq:vertexTaylor}), extracted as described in the text. The last
entry (EGM) is the chiral N$^3$LO interaction of Epelbaum, Gl\"ockle, and Mei{\ss}ner \cite{EpelbaumGlockleMeissner2005} at
Lippmann-Schwinger cutoff $450$ and spectral-function cutoff $500~\mathrm{MeV}$, the pair with the lowest
cutoffs among the four published there. For AV18, CD Bonn, and N$^3$LO 500 the calculated $S$-wave
ANCs reproduce the published values \cite{Wiringa1995,Machleidt2001,EntemMachleidt2003} to four digits.}
\label{tab:potentials}
\begin{ruledtabular}
\begin{tabular}{lccccc}
Interaction & $B_d$ & $P_D$ & $f_D^{(V)}$ & $\langle T_{np}\rangle_V$ & $c$ \\
 & (MeV) & (\%) & (\%) & (MeV) & (fm$^{2}$) \\
\hline
AV18 \cite{Wiringa1995}            & 2.2246 & 5.760 & 39.348 & 217.9 & 0.5776 \\
CD Bonn \cite{Machleidt2001}       & 2.2233 & 4.854 & 32.653 & 124.7 & 0.5644 \\
N$^3$LO 500 \cite{EntemMachleidt2003} & 2.2246 & 4.514 & 22.693 &  84.0 & 0.5422 \\
N$^3$LO 600 \cite{EntemMachleidt2003} & 2.2246 & 4.427 & 24.147 & 104.3 & 0.5446 \\
N$^3$LO EGM 450/500                & 2.2164 & 2.727 &  9.896 &  63.1 & 0.4933 \\
\end{tabular}
\end{ruledtabular}
\end{table}

The neutron-proton interaction term in a post-form transfer amplitude contains the
form factor $V_{np}\phi_d$ \cite{ButtleGoldfarb1964}. The deuteron is the coupled
$^3S_1$-$^3D_1$ bound state. In coordinate space its components are the radial functions
$u_l(r)$, with $u_S\equiv u_0$ and $u_D\equiv u_2$ normalized as
$\int_0^\infty dr\,(u_S^2+u_D^2)=1$; in momentum space they are
\begin{equation}
\tilde\phi_l(p)=(-i)^l\sqrt{\frac{2}{\pi}}\int_0^\infty dr\,r\,j_l(pr)\,u_l(r),
\label{eq:fourierbessel}
\end{equation}
with $p$ the neutron-proton relative momentum and $j_l$ a spherical Bessel function, which carry
the same normalization with the measure $p^2dp$, with
$\tilde\phi_S\equiv\tilde\phi_0$ and $\tilde\phi_D\equiv\tilde\phi_2$. The Fourier phase $(-i)^l$ is
included in $\tilde\phi_l$. The partial-wave components of the form factor
$\langle p|V_{np}|\phi_d\rangle$ are
\begin{equation}
\tilde D_l(p)=\sum_{l'}\int_0^\infty dp'\,p'^2
V_{ll'}(p,p')\tilde\phi_{l'}(p').
\label{eq:vertexdef}
\end{equation}
I call $\tilde D_S\equiv\tilde D_0$ and $\tilde D_D\equiv\tilde D_2$ the $S$- and $D$-wave
form factors. Through the tensor couplings, $\tilde D_S$ receives a contribution $V_{02}\tilde\phi_D$
and $\tilde D_D$ a contribution $V_{20}\tilde\phi_S$.
The bound-state equation is
\begin{equation}
\frac{p^2}{m_2}\tilde\phi_l(p)+\tilde D_l(p)
=-B_d\tilde\phi_l(p),
\label{eq:momspace}
\end{equation}
where $m_2=2\mu c^2/(\hbar c)=4.758~\mathrm{fm^{-1}}$ and $\mu$ is the
neutron-proton reduced mass. Momenta are in $\mathrm{fm^{-1}}$, and $B_d$ in this
equation is an energy in $\mathrm{fm^{-1}}$; multiplication by $\hbar c$ converts it
to MeV. The conversion factor is $\hbar c/m_2=41.471~\mathrm{MeV\,fm^2}$.
The kinetic operator is diagonal in $l$, while the tensor interaction enters
through $\tilde D_l$. Rearranging gives the exact channel-by-channel identity
\begin{equation}
\tilde D_l(p)=-\frac{p^2+\kappa^2}{m_2}\tilde\phi_l(p),
\qquad \kappa^2=m_2B_d,
\label{eq:identity}
\end{equation}
the momentum-space form of the relation $V_{np}\phi_d=-(T+B_d)\phi_d$ used in
Ref.~\cite{ButtleGoldfarb1964}. It holds for nonlocal interactions as well, so the form
factor follows from the wave function alone.

\begin{figure*}[!t]
\includegraphics[width=\textwidth]{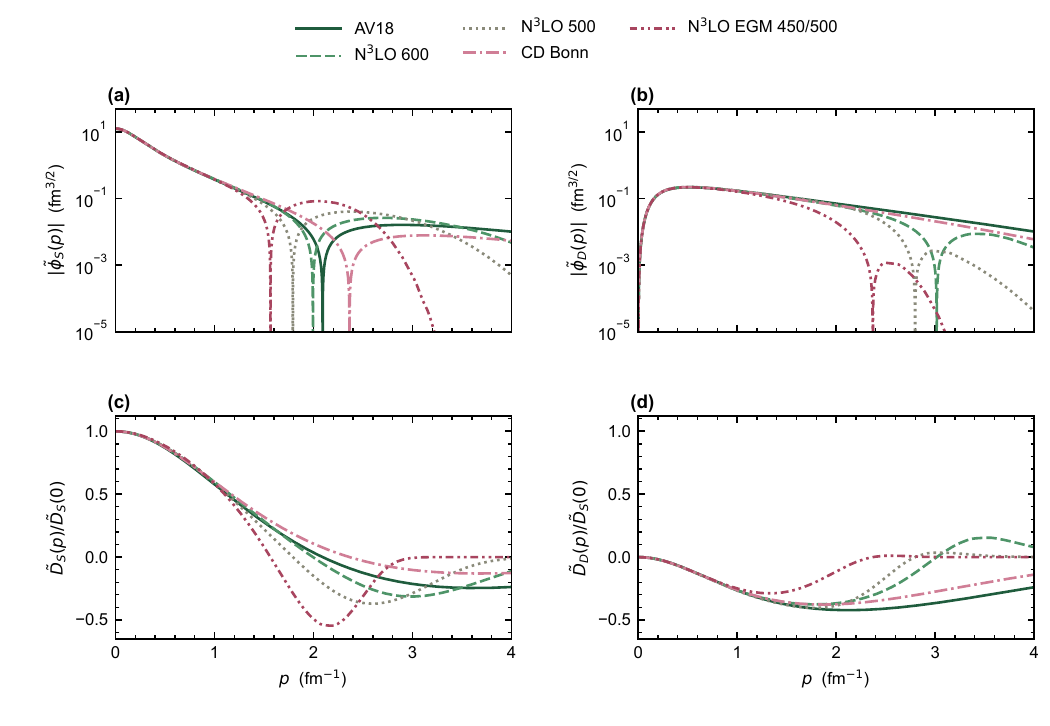}
\caption{Deuteron wave functions and form factors for the five interactions
in Table~\ref{tab:potentials}.
(a), (b) Magnitudes of the $S$- and $D$-wave components of the normalized wave function,
with logarithmic vertical scales. (c), (d) Signed form factors, both divided by $\tilde D_S(0)$ for each
interaction and shown on the same vertical scale. Signs follow Eq.~(\ref{eq:fourierbessel}).}
\label{fig:structure}
\end{figure*}

Figure~\ref{fig:structure} shows the deuteron wave function in the upper row and the form
factor in the lower row, with the $S$ component on the left and the $D$ component on the right;
each line is one interaction of Table~\ref{tab:potentials}. Panels (a) and (b) give
$|\tilde\phi_S|$ and $|\tilde\phi_D|$ on logarithmic scales, so a node of a component
appears as a downward spike. The lower panels are described after
Eq.~(\ref{eq:vertexTaylor}), where the normalized form factor is defined.

Outside the range of the interaction, $u_S(r)\to A_S e^{-\kappa r}$, with $A_S$ the
$S$-wave ANC. With Eq.~(\ref{eq:fourierbessel}) the momentum-space component can be written
\begin{equation}
\tilde\phi_S(p)=\sqrt{\frac{2}{\pi}}
\left[\frac{A_S}{p^2+\kappa^2}+R(p^2)\right].
\label{eq:polesplit}
\end{equation}
This decomposition is exact and neglects nothing. The first term is the Fourier transform of
the pure exponential $A_Se^{-\kappa r}$ continued to the origin; the remainder $R$ is the
transform of the difference $u_S-A_Se^{-\kappa r}$, which contains the whole interior of the
wave function. That difference falls off faster than $e^{-\kappa r}$, so $R$ has no
singularity at $p^2=-\kappa^2$: it is finite and analytic there. For a local interaction,
a one-pion tail produces a more distant singularity near $p^2=-(\kappa+m_\pi)^2$, with $m_\pi$ the
pion mass in inverse-length units
(Sec.~\ref{sec:parameter}). The asymptotic tail therefore appears in momentum space as the singular term at
$p^2=-\kappa^2$, which dominates at small $p$, while the interior enters through $R$, whose
relative weight grows with $p$; at large $p$ it cancels the $1/p^2$ falloff of the asymptotic term.
Defining $C=-\sqrt{2/\pi}/m_2$, Eq.~(\ref{eq:identity}) gives
\begin{equation}
\tilde D_S(p)=C\left[A_S+(p^2+\kappa^2)R(p^2)\right].
\label{eq:vertexexpansion}
\end{equation}
The singularity of the wave function at $p^2=-\kappa^2$ has canceled, and the form factor
continued to that point equals $CA_S$. Away from it, the correction depends on the
interaction. Whether the ANC dominates an amplitude depends on this correction over the
momenta the reaction samples. The same separation underlies descriptions of peripheral halo
observables in terms of asymptotic structure \cite{Capel2018}.

The $D$ component obeys the same identity but vanishes as $p^2$ at the origin. Separating its
asymptotic term in the same way leaves a regular interior term in its low-momentum slope, so the
$D$-state ANC does not by itself fix that slope.

The $D$ wave matters more for the form factor than its probability suggests. The
probability $P_D=\int dp\,p^2|\tilde\phi_D|^2$ measures the $D$ wave in the wave function.
By Eq.~(\ref{eq:identity}), the $D$-wave fraction $f_D^{(V)}$ of the matrix element
$\langle\phi_d|V_{np}|\phi_d\rangle$ is
\begin{equation}
f_D^{(V)}=
\frac{\int dp\,p^2(p^2+\kappa^2)|\tilde\phi_D|^2}
{\int dp\,p^2(p^2+\kappa^2)(|\tilde\phi_S|^2+|\tilde\phi_D|^2)}.
\label{eq:pdbar}
\end{equation}
The $D$-wave contribution includes both the diagonal $V_{DD}$ term and the tensor
coupling $V_{DS}$ from the $S$ component [Eq.~(\ref{eq:vertexdef})].
The factor $p^2+\kappa^2$ weights each component toward high momentum, where the $D$ wave is
relatively larger [Figs.~\ref{fig:structure}(a) and \ref{fig:structure}(b)], so $f_D^{(V)}$ is 3.6 to 6.8 times $P_D$ for the five interactions of
Table~\ref{tab:potentials}. This is the enhancement discussed in Refs.~\cite{Bailey2016,Bailey2017},
where $f_D^{(V)}$ is introduced as the $D$-state fraction of $\langle\phi_d|V_{np}|\phi_d\rangle$.
Equation~(\ref{eq:pdbar}) is its momentum-space form. Evaluated with the present wave functions it
gives 39.35 percent for AV18 and 32.65 percent for CD Bonn (Table~\ref{tab:potentials}),
within 0.06 percentage point of the values 39.4 and 32.6 percent of Ref.~\cite{Bailey2016}. Both $P_D$ and $f_D^{(V)}$ depend on the interaction
representation. In the following sections I average the $S$-wave form factor with
several weights (Sec.~\ref{sec:where}). The two interaction-dependent weights contain both
bound-state components. The averaged form factor is $\tilde D_S$, which already includes the
tensor coupling to $\tilde\phi_D$; no $|\tilde D_D|$ term is added to it. In a transfer amplitude
the two components enter coherently, each multiplied by its own spin-angle function and by
the distorted waves, so $S$-$D$ cross terms can survive when the distortion or a
polarization selects a direction. Summing the squared form factor over spin projections and
integrating over the direction of $\bm p$ removes them by orthogonality of the spin-angle
functions and leaves a quantity proportional to $|\tilde D_S|^2+|\tilde D_D|^2$, which keeps
the size of the $D$ component and discards its phase relative to the $S$ component.

To compare the shapes of the $S$-wave form factors, divide each by its own value at zero
momentum; the initial slope of that ratio defines the finite-range coefficient $c_i$.
With $x=p^2$,
\begin{equation}
\begin{split}
F_i(x)&\equiv\frac{\tilde D_{S,i}(x)}{\tilde D_{S,i}(0)}
=1-c_i x+\mathcal O(x^2),\\
c_i&\equiv-\frac{\tilde D'_{S,i}(0)}{\tilde D_{S,i}(0)}.
\end{split}
\label{eq:vertexTaylor}
\end{equation}
The index $i$ labels the interactions, and primes denote derivatives with respect to $x$.
Figure~\ref{fig:structure}(c) shows $F_i$ for the five interactions, and
Fig.~\ref{fig:structure}(d) shows $\tilde D_D$ divided by the same $\tilde D_S(0)$, so the two
panels share one scale. By Eq.~(\ref{eq:identity}) each form factor is the corresponding
wave-function component multiplied by $-(p^2+\kappa^2)/m_2$: the factor shifts weight toward
higher momenta and leaves the nodes and the ratio of the $D$ and $S$ components at each
momentum unchanged. The curves in panel (c) have similar shapes near zero momentum.
For the five interactions studied here, $\tilde D_S(0)\neq0$ and $c_i>0$, so $c_i p^2$ is
the leading fractional decrease of the form factor from its zero-momentum value; the values
are in Table~\ref{tab:potentials}. Equation~(\ref{eq:vertexTaylor}) is the finite-range expansion
of Ref.~\cite{ButtleGoldfarb1964}. Reference~\cite{Bailey2016} tabulates $c_i^{-1/2}$, as
$\kappa$: 1.32 and $1.33~\mathrm{fm^{-1}}$ for AV18 and CD Bonn, against 1.316 and
$1.331~\mathrm{fm^{-1}}$ here.

The adiabatic distorted-wave approximation makes two approximations to the $n+p+A$ problem with
Hamiltonian $T_R+T+V_{np}+U_{nA}+U_{pA}$, with $T_R$ the kinetic energy in the deuteron-target
separation $\mathbf R$ and $\Psi$ its three-body wave function. First, Johnson and Soper replace the neutron-proton
Hamiltonian by $-B_d$, assuming breakup excitation energies small compared with the beam energy
\cite{JohnsonSoper1970}. The neutron-proton separation then enters only as a parameter, breakup is
included to all orders within this assumption, and at zero separation the deuteron moves in
$U_{nA}+U_{pA}$, conventionally with each nucleon potential evaluated at half the deuteron
energy. Second, for a finite-range
$V_{np}$, Johnson and Tandy expand $\Psi$ in Weinberg states, the solutions of
$(T+\alpha_iV_{np})\phi_i=-B_d\phi_i$ at fixed energy, with $\alpha_i$ the
interaction-strength eigenvalues, which are complete within the range of
$V_{np}$, and keep the first component, $\phi_1=\phi_d$ \cite{JohnsonTandy1974,Laid1993}. Its
amplitude $\chi(\mathbf R)\propto\langle\phi_d|V_{np}|\Psi\rangle$ is not the elastic projection
$\langle\phi_d|\Psi\rangle$ of the distorted-wave Born approximation. Breakup states $\phi_k$ are
orthogonal to $\phi_d$ but not with weight $V_{np}$, since
$\langle\phi_d|V_{np}|\phi_k\rangle=-\langle\phi_d|T|\phi_k\rangle$, so $\chi$ carries the breakup
amplitude within the range of $V_{np}$. Neglecting couplings to higher Weinberg components gives
a single-channel approximation to $\chi$; for local nucleon-target potentials its potential is
\begin{equation}
\begin{split}
U_{dA}(\mathbf R)&=\big\langle\bar\phi_1\big|U_{nA}(\mathbf R-\tfrac{\mathbf r}{2})
+U_{pA}(\mathbf R+\tfrac{\mathbf r}{2})\big|\phi_d\big\rangle,\\
\bar\phi_1&=\frac{V_{np}\phi_d}{\langle\phi_d|V_{np}|\phi_d\rangle}.
\end{split}
\label{eq:johnsontandy}
\end{equation}
For $^{132}$Sn$(d,p)$ at deuteron energies of 30 and 100 MeV, Pang \textit{et al.} found the
forward-angle transfer amplitude to be dominated by the first Weinberg component extracted from
continuum-discretized calculations \cite{Pang2013}. The neutron-proton
interaction therefore enters an adiabatic calculation twice: through the form factor in the
transition operator, and through the weight $\bar\phi_1\propto D$ in $U_{dA}$. Continuum-discretized
calculations keep the breakup energies that the adiabatic approximation freezes, and the Faddeev
calculation of Ref.~\cite{Deltuva2018} solves the three-body problem without either
approximation.

Both approximations act on $\Psi$. In the post-form transition operator $V_{np}+U_{pA}-U_{pB}$,
with $U_{pB}$ the proton optical potential for the residual nucleus $B$, the $V_{np}$ term needs
$\Psi$ only within the range of $V_{np}$. The remnant term $U_{pA}-U_{pB}$
samples $\Psi$ beyond the range of $V_{np}$, so its dependence on the neutron-proton interaction
does not pass through $V_{np}\phi_d$; adiabatic calculations usually omit it, and so does the
analysis here. It is expected to be small for heavy targets. In light systems its effect at the
peak of the angular distribution is below 5 percent for $^{12}$C(d,p)$^{13}$C(g.s.) above 8 MeV,
but about 10 percent for the weakly bound $1/2^+$ state of $^{13}$C at 3.09 MeV, growing as the
$s_{1/2}$ neutron binding decreases \cite{Delaunay2005}, and 27 percent
for $^{11}$Be(p,d) at 35 MeV, where the adiabatic approximation also departs from the Faddeev
result \cite{NunesDeltuva2011}; in $^{16}$O(d,p) it reaches 30 percent at large angles
\cite{Shubhchintak2019}.

The structure integral through which the nonlocal ADWA energy shift depends on $V_{np}$ is
the weighted kinetic energy $\langle T_{np}\rangle_V$ of Eq.~(\ref{eq:tnpdef}). Reduced with the bound-state equation $V_{np}\phi_d=-(T+B_d)\phi_d$,
\begin{equation}
\langle T_{np}\rangle_V
=\frac{\langle\phi_d|V_{np}T|\phi_d\rangle}{\langle\phi_d|V_{np}|\phi_d\rangle}
=\langle T\rangle+\frac{\langle T^2\rangle-\langle T\rangle^2}{\langle T\rangle+B_d},
\label{eq:tnpvariance}
\end{equation}
where the expectation values on the right are taken in the normalized deuteron state. It is
the ratio used in Refs.~\cite{TimofeyukJohnson2013PRL,Bailey2017}. In momentum space it equals
$(\hbar c/m_2)\langle p^2\rangle$, with $\langle p^2\rangle$ the mean squared momentum under the weight
$(p^2+\kappa^2)\rho(p)$ and $\rho$ the deuteron momentum density summed over the $S$ and $D$
components; Sec.~\ref{sec:where} uses this as the ADWA weight.
For AV18 the result is $217.9~\mathrm{MeV}$, compared with
$218.0~\mathrm{MeV}$ in Ref.~\cite{Bailey2017}; the agreement checks the structure integral
alone, without the entrance distorted wave or the optical-potential response. For CD Bonn the
value here, 124.7 MeV, is 11 percent above the 112.5 MeV of Ref.~\cite{Bailey2017}. Its weighted
kinetic energy still gains 12 MeV above 5 fm$^{-1}$: the contribution
from $p<5$ fm$^{-1}$ is 112.3 MeV (113.1 MeV with the truncated weight renormalized), while $P_D$ and
the $D$-state share of $\langle\phi_d|V_{np}|\phi_d\rangle$ barely change, so the difference is consistent with a
momentum truncation near 1 GeV$/c$. For AV18, of the 217.9 MeV,
the mean kinetic energy $\langle T\rangle$ contributes 19.8 MeV and the variance term 198.1 MeV. For CD Bonn and EGM 450/500
the variance term is 109.1 of 124.7 MeV and 50.1 of 63.1 MeV. The weighted kinetic energy is
dominated by the spread of the kinetic energy, which the high-momentum tail of the
deuteron controls.

\begin{table}[t]
\caption{The finite-range coefficient and $\langle T_{np}\rangle_V$ along the unitary family
of Eq.~(\ref{eq:unitary}), starting from AV18. The third column gives the change $\Delta c_\theta=c_\theta-c_0$ of $c$ in
units of the realistic range $\Delta c=0.0843~\mathrm{fm^2}$; the fifth column gives the change
of $\langle T_{np}\rangle_V$ in units of its realistic range, 154.8 MeV (63.1 to 217.9 MeV in
Table~\ref{tab:potentials}). Binding energy, asymptotic normalizations, $P_D$, and the coupled
scattering matrix are the same in every row. At $\theta=0$ the values are those of AV18.}
\label{tab:tnp}
\begin{ruledtabular}
\begin{tabular}{ccccc}
$R_h$ & $\theta$ & $\Delta c_\theta/\Delta c$ & $\langle T_{np}\rangle_V$ & change \\
(fm) & & & (MeV) & $/154.8$ \\
\hline
1.5 & 0.05 & 0.011 & 314.0 & 0.62 \\
1.5 & 0.10 & 0.032 & 518.8 & 1.94 \\
2.0 & 0.05 & 0.067 & 307.0 & 0.58 \\
2.0 & 0.10 & 0.154 & 454.8 & 1.53 \\
2.0 & 0.40 & 1.079 & 1377.2 & 7.49 \\
3.0 & 0.10 & 0.504 & 284.7 & 0.43 \\
3.0 & 0.40 & 3.264 & 644.9 & 2.76 \\
\end{tabular}
\end{ruledtabular}
\end{table}

The same structure quantities also admit coordinate-space expressions. By Eq.~(\ref{eq:identity}),
$D=-(T+B_d)\phi_d$ can be evaluated from local derivatives of the bound state even for a
nonlocal interaction, and its low moments are
given in Eq.~(\ref{eq:coordmoments}). For the structure integral, $T$ is Hermitian, so
$\langle T^2\rangle=\|T\phi_d\|^2$ (the regularity $u_S\sim r$, $u_D\sim r^3$ at the origin and
the decaying tails remove boundary terms), and with $\int dr\,(|u_S|^2+|u_D|^2)=1$,
\begin{equation}
\langle T^2\rangle=\left(\frac{\hbar c}{m_2}\right)^2\int_0^\infty dr
\left[\,|u_S''|^2+\left|u_D''-\frac{6u_D}{r^2}\right|^2\right].
\label{eq:curvature}
\end{equation}
Together with Eq.~(\ref{eq:tnpvariance}), Eq.~(\ref{eq:curvature}) gives an exact
coordinate-space expression for the weighted kinetic energy that
Ref.~\cite{TimofeyukJohnson2013PRL} associates with the confinement of the
neutron-proton motion within the range of $V_{np}$. Where that curvature sits depends on the interaction: for
AV18, 96 percent of $\langle T^2\rangle$ comes from $r<1.5$ fm and 99 percent from $r<2$ fm,
while for EGM 450/500 the shares are 14 and 42 percent, the regulator spreading the curvature
to larger $r$ as it spreads $D(r)$.

The calculations are done in momentum space chiefly because the kinetic operator is diagonal
there. Equations~(\ref{eq:identity}) and (\ref{eq:tnpvariance}) then reduce the form factor and
the kinetic-energy averages to pointwise multiplication and momentum integration, without
differentiating the wave function. The interaction matrices used here are also supplied in
momentum space, and Eq.~(\ref{eq:polesplit}) separates the $S$-wave asymptotic term analytically.

\section{Representation dependence at fixed two-body observables}
\label{sec:representation}

The interactions in Table~\ref{tab:potentials} are independent fits. Their differences
can reflect both physical input and representation. A family of
scattering-equivalent interactions holds two-body observables fixed and isolates
the effect of changing representation
\cite{PolyzouGlockle1990}.
Let $h$ be a normalized radial function with compact support $0<r<R_h$,
chosen as $h(r)\propto\exp[-1/(1-x^2)]$ with $x=2r/R_h-1$.
It and all its derivatives vanish at both endpoints.
Orthogonalizing $h(r)\cos(2\pi r/R_h)$ against $h$ gives a second
normalized function $h'$. A rotation in the plane they span is
\begin{equation}
\begin{split}
U(\theta)=1&+(\cos\theta-1)
\left(|h\rangle\langle h|+|h'\rangle\langle h'|\right)\\
&+\sin\theta\left(|h\rangle\langle h'|-|h'\rangle\langle h|\right).
\end{split}
\label{eq:unitary}
\end{equation}
The generators lie entirely in the $S$ component of the deuteron sector.
The rotation changes the coupled $^3S_1$-$^3D_1$ interaction through
$H_\theta=UHU^\dagger$, while acting as the identity in other sectors; with $H=T+V_{np}$ the
transformed neutron-proton interaction is $V_\theta=H_\theta-T$.
Compact support leaves the asymptotic wave functions unchanged, so the
bound-state energy, both asymptotic normalizations, and the coupled scattering matrix
are preserved.

For the family in this section and the folding example in
Sec.~\ref{sec:recovery}, the starting interaction is AV18 on a
400-point momentum mesh. The default parameters are
$R_h=2.0~\mathrm{fm}$ and $\theta=0,0.05,\ldots,0.40$.
The binding energy, the ANCs, both eigenphases, and the mixing parameter are preserved to
numerical precision. Applying $U$ to the original
eigenvector produces the transformed state without further diagonalization;
independent diagonalization of $H_\theta$ checks that construction.

Table~\ref{tab:tnp} follows the two quantities that Sec.~\ref{sec:identity} separates: the
finite-range coefficient, which sets the low-momentum shape of the form factor, and
$\langle T_{np}\rangle_V$, which sets the ADWA energy shift. Changes of $c$ are measured
relative to $\Delta c$, the range of the coefficients across the five interactions of
Table~\ref{tab:potentials}. A rotation confined to
$r<1.5$ fm with $\theta=0.05$ moves $c$ by one percent of its realistic range and
$\langle T_{np}\rangle_V$ from 217.9 to 314.0 MeV, a change equal to 62 percent of the range
spanned by the five interactions. At $\theta=0.10$ the average reaches 518.8 MeV, beyond every realistic value,
while $c$ has moved by three percent of its range. The increase comes mainly from momenta above
$6~\mathrm{fm^{-1}}$, and at $R_h=1.5$ fm part of it
lies beyond the $60~\mathrm{fm^{-1}}$ momentum mesh; the values in Table~\ref{tab:tnp} include
that part, evaluated from the analytic transforms of the two functions spanning the rotation. The rotation range and angle are not set by
the error scale of any reaction approximation, so reaching 62 percent of the realistic span does
not mean accounting for 62 percent of the realistic differences. The ratio
of the two responses grows as the support shrinks. By Eq.~(\ref{eq:tnpvariance}) the average is dominated by the kinetic-energy variance, which
for AV18 comes almost entirely from the interior, Eq.~(\ref{eq:curvature}); a transformation
confined to $r<R_h$ reshapes the interior derivatives directly, while $c$ is the signed
mean square radius of the $S$-wave form factor divided by six, and the tail beyond $R_h$ is
untouched. Every row has the same binding energy, asymptotic normalizations,
$D$-state probability, and coupled scattering matrix.

\begin{figure*}[!t]
\includegraphics[width=\textwidth]{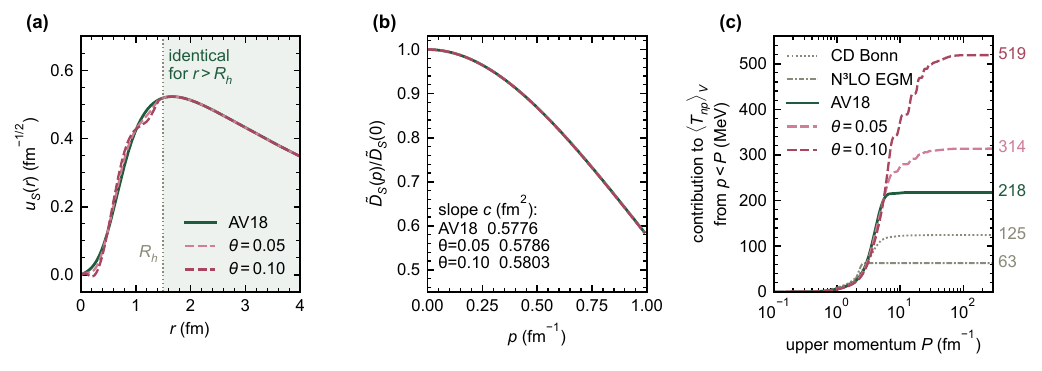}
\caption{One deuteron in three representations: AV18 (solid) and two members of the unitary
family of Eq.~(\ref{eq:unitary}) with $R_h=1.5$ fm, $\theta=0.05$ and 0.10 (dashed), which share
the binding energy, both ANCs, and the coupled scattering matrix. (a) $S$-wave radial function;
the dotted line marks the support radius $R_h$ of the rotation, and in the shaded region $r>R_h$
the three functions coincide. (b) Normalized $S$-wave form factor at low momentum, with the
finite-range coefficients $c$. (c) Contribution to $\langle T_{np}\rangle_V$ of
Eq.~(\ref{eq:tnpdef}) from momenta $p<P$, plotted against the upper limit $P$; the right end of
each curve is the full value; CD Bonn (dotted) and EGM 450/500 (dash-dotted) are shown in gray. The realistic
curves level off by $P\approx40~\mathrm{fm^{-1}}$; the family curves keep rising beyond it, carried
by the slowly decaying high-momentum components that the rotation adds. The numbers at the right
are the limits $P\to\infty$.}
\label{fig:mechanism}
\end{figure*}

Figure~\ref{fig:mechanism} shows the mechanism for $R_h=1.5$ fm. In each panel the solid line is
AV18 and the dashed lines are the members with $\theta=0.05$ and 0.10. Panel (a) gives the
$S$-wave radial function $u_S(r)$; the shaded region $r>R_h$ is where the three coincide.
Panel (b) gives the normalized form factor $F(p)$ of Eq.~(\ref{eq:vertexTaylor}) with the slope
$c$ of each. Panel (c) gives the contribution to $\langle T_{np}\rangle_V$ from momenta $p<P$,
as a function of the upper limit $P$, with CD Bonn and EGM 450/500 (gray) for scale; the right
end of each curve is the full value.

Each value uses the prescription that an ADWA calculation in the new representation would
use: the weighting is rebuilt from the
transformed interaction $V_\theta$ and the kinetic operator is untransformed.
With $|\phi_\theta\rangle=U|\phi_d\rangle$, the form factor obeys
\begin{equation}
V_\theta|\phi_\theta\rangle
=U(V_{np}|\phi_d\rangle)-[T,U]|\phi_d\rangle.
\label{eq:formfactortransform}
\end{equation}
The commutator term accounts for the difference between rebuilding the form factor
with the original kinetic operator $T$ and transporting it with $U$. Transforming both
the weighting operator, to $UV_{np}U^\dagger$ of Eq.~(\ref{eq:operatortransform}), and the
kinetic operator, to $UTU^\dagger$, restores 217.9 MeV by unitarity, as the folded kernel is
restored in Sec.~\ref{sec:recovery}; transforming the kinetic operator alone does not.

Timofeyuk and Johnson trace the size of the shift to the confinement of the neutron-proton
relative motion within the range of $V_{np}$ \cite{TimofeyukJohnson2013PRL}. In their
derivation the nucleon nonlocality displaces each nucleon by distances of order its range, which is
comparable to the range of $V_{np}$; to second order in that displacement the neutron-proton
wave function enters through its curvature, so the nonlocality acts as a dependence on nucleon
momentum that the internal motion of the deuteron supplies, weighted by $V_{np}\phi_d$. In the
local-equivalent description this dependence appears as a shift of the energy at which the
nucleon potentials are evaluated. Table~\ref{tab:tnp}
shows that this weighted kinetic energy is not fixed by two-body observables, so a reaction
prescription that depends on it inherits the freedom of representation.

The entries describe the structure integral in the energy-shift prescription. They do not
give the energy shift of a consistently transformed nonlocal calculation: the transformed
nucleon-target interactions, $U_{nA}+U_{pA}+W_3$ in Eq.~(\ref{eq:embedded}), need not keep the
Perey-Buck form assumed in deriving $\Delta E=\langle T_{np}\rangle_V/2$
\cite{TimofeyukJohnson2013PRL}, so half of a tabulated value is not that shift. The
folded-kernel residuals of Sec.~\ref{sec:recovery}, computed for a local real potential, are
a separate diagnostic and do not measure the nonlocal case either.

The same family also reaches the realistic variation in the low-momentum coefficients. The
five realistic interactions occupy $c_i=0.49334$ to $0.57761~\mathrm{fm^2}$. The default
family changes $c_\theta$ from $0.57761$ to $0.66852~\mathrm{fm^2}$.
The two widths are comparable, but their intervals lie on opposite sides of the
AV18 value. The family matches the size of the model variation in the low-momentum
coefficient, not its absolute interval.
The rotation also changes the zero-momentum form factor relative to the fixed ANC
normalization. Their ratio, $q_i\equiv\tilde D_{S,i}(0)/(C_iA_{S,i})$, with $C_i=-\sqrt{2/\pi}/m_{2,i}$, equals one when
the two normalizations coincide. With $C A_S$ held fixed, $q$
changes from $0.9693$ to $0.9646$, a shift of 0.469 percent, against a
range of 0.457 percent among the five interactions. Both the offset and the finite-range
term of the low-momentum expansion, Eq.~(\ref{eq:masteranc}) of Sec.~\ref{sec:parameter}, can change by amounts comparable to the
model variation without any change of two-body observables.

The realistic interactions differ by 0.37 percent in $B_d$ and 0.34 percent
in $A_S$. Normalizing each form factor at zero removes its multiplicative factor,
but their remaining shape differences can still include differences of fits.
The unitary family shows that changes of the size found among the realistic interactions can
arise from representation alone; it does not divide the differences among the independent fits
into physical and representational parts.

\section{How the integral enters the adiabatic calculation}
\label{sec:whyadwa}

Two features of the adiabatic projection expose the high-momentum components that the
transformation changes; continuum-discretized and Faddeev calculations treat both differently.

The first is the weight. A Perey-Buck nucleon potential couples the wave function at different
nucleon-target separations, with a Gaussian weight of range $\beta$. Inside the deuteron this
displaces the neutron-proton separation, by $\alpha$ times the nucleon displacement with
$\alpha=A/(A+1)$ and $A$ the target mass number, so the wave function is compared with a displaced
copy of itself. The Gaussian-averaged overlap of the weight $\bar\phi_1$ of
Eq.~(\ref{eq:johnsontandy}) with the displaced $\phi_d$ is the coefficient $M_0$ of Timofeyuk and
Johnson, which multiplies the sum of the nucleon potentials in the leading-order local-equivalent
adiabatic equation \cite{TimofeyukJohnson2013PRL,TimofeyukJohnson2013,TimofeyukJohnson2020}. The
averaging suppresses high internal momenta: in momentum space each component is multiplied by
$e^{-\alpha^2\beta^2p^2/4}$, whose e-folding scale $2/(\alpha\beta)$ is about
$2.4~\mathrm{fm^{-1}}$ for $\beta=0.85$ fm, and
\begin{equation}
M_0=\big\langle e^{-\alpha^2\beta^2p^2/4}\big\rangle_{\rm ADWA},
\label{eq:m0}
\end{equation}
the average under the weight $(p^2+\kappa^2)\rho(p)$ that defines $\langle p^2\rangle$ in
Sec.~\ref{sec:identity}. In the continuum-discretized calculation the corresponding coefficient
of the elastic channel is the overlap of $\phi_d$ with its displaced copy, which contains no
$V_{np}$ \cite{GomezRamosTimofeyuk2018}; it is Eq.~(\ref{eq:m0}) with $\rho(p)$ alone. That
average is dominated by large neutron-proton separations, where the interactions have nearly
identical asymptotic tails, whereas the factor $p^2+\kappa^2$ moves the ADWA average to the separations
where $V_{np}$ acts and the interactions differ. For $\beta=0.85$ fm the range of the overlap
across the five interactions is 27 percent of their mean with the ADWA weight and 1.3 percent
with $\rho$ alone.

The second is the frozen neutron-proton energy. ADWA replaces the neutron-proton Hamiltonian by
$-B_d$, assigning the deuteron energy to the internal propagation instead of the excitation
energies of the continuum, while the momentum dependence of the nonlocal potentials reads the
internal motion sampled by $V_{np}\phi_d$ as nucleon energy. This is the origin of the
evaluation energy $E_d/2+\Delta E$. In the Faddeev resolvent the internal kinetic energy is
retained, and energy conservation limits the nucleon-target energy to about 9 MeV for
$^{26}$Al$(d,p)$ at 12 MeV, against shifted energies of 50 to 72 MeV \cite{Deltuva2018}. A
continuum-discretized calculation retains the excitation energies of its continuum bins, and
the small sensitivity it reports grows with beam energy, an increase associated with the
higher-energy bins \cite{GomezRamosTimofeyuk2018}.

For a short nonlocality range the lost overlap is proportional to the mean squared momentum,
$1-M_0\simeq\alpha^2\beta^2\langle p^2\rangle/4$, and $\langle p^2\rangle$ is
$\langle T_{np}\rangle_V$ in units of $\hbar c/m_2$. The same average, once the
local-equivalent equation is solved self-consistently in energy, gives the leading energy-shift prescription $\Delta E=\langle T_{np}\rangle_V/2$
\cite{TimofeyukJohnson2013PRL}. Converting the full lost overlap into energy
units defines a bounded counterpart of the weighted kinetic energy,
\begin{equation}
T_\beta=\frac{4}{\alpha^2\beta^2}\,\frac{\hbar c}{m_2}\,(1-M_0),
\label{eq:tbeta}
\end{equation}
which tends to $\langle T_{np}\rangle_V$ as $\beta\to0$. The two averages count low momenta
alike. $\langle T_{np}\rangle_V$ counts every momentum with its full $p^2$, while in $T_\beta$
the contribution of each momentum approaches a constant well above $2/(\alpha\beta)$. A rotation confined to short distances changes mostly the momenta
beyond that scale, so expanding the nonlocality to lowest order, which keeps the full $p^2$,
strongly amplifies the response of the overlap: at $R_h=1.5$ fm and
$\theta=0.05$ the change is 96 MeV in $\langle T_{np}\rangle_V$ and 6 MeV in $T_\beta$, which
spans 47.3 to 94.6 MeV among the five interactions; the change is 12 percent of that span against
62 percent for $\langle T_{np}\rangle_V$, and the reduction is similar for $\beta$ between 0.7 and
1.0 fm. A wider rotation reaches lower momenta, and the two then respond alike: at $R_h=3.0$ fm and
$\theta=0.10$ they move by 0.43 and 0.37 of their realistic spans. With the Gaussian factor kept, the overlap $M_0$ thus still depends on the representation, but
responds far less than its small-$\beta$ expansion when the rotation is confined to short
distances, and the expansion also widens the spread among the realistic interactions. A
prescription that evaluates the nucleon potentials at $E_d/2+\langle T_{np}\rangle_V/2$
\cite{JohnsonTimofeyuk2014} uses the amplified quantity directly. The continuum-discretized
elastic coefficient, built with $\rho$ alone, is nearly untouched by the rotation at $R_h=1.5$ fm and $\theta=0.05$.

\begin{figure}[t]
\includegraphics[width=\columnwidth]{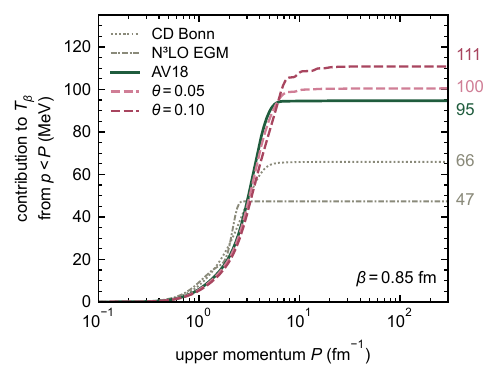}
\caption{Contribution to $T_\beta$ of Eq.~(\ref{eq:tbeta}) from momenta $p<P$, for
$\beta=0.85$ fm and $A=40$, with the lines of Fig.~\ref{fig:mechanism}(c): AV18 (solid), the
family members $\theta=0.05$ and 0.10 at $R_h=1.5$ fm (dashed), CD Bonn and EGM 450/500 (gray).
The numbers at the right are the limits $P\to\infty$.}
\label{fig:tbeta}
\end{figure}

Figure~\ref{fig:tbeta} shows the same accumulation as Fig.~\ref{fig:mechanism}(c) with $p^2$
replaced by the bounded factor of Eq.~(\ref{eq:tbeta}), so each curve ends at $T_\beta$. Up to
about $1.5~\mathrm{fm^{-1}}$ the two accumulations agree within 10 percent; beyond it the bounded
factor grows more and more slowly, and the large high-momentum contribution that the rotation adds to
$\langle T_{np}\rangle_V$ shrinks to a small step.

\section{The low-momentum expansion}
\label{sec:parameter}

The finite-range expansion of Ref.~\cite{ButtleGoldfarb1964} explains why the low-momentum
averages of the form factor differ little among the interactions, and why the transformation of
Sec.~\ref{sec:representation} reaches them at leading order through the finite-range coefficient
and the offset $1-q$, which a small rotation leaves nearly unchanged. For the Hulth\'en deuteron the form
factor is a Yukawa function with a range of about 0.6 fm; $c$ is the square of that range
and $q=1-\kappa^2c$ \cite{ButtleGoldfarb1964}.

A factorized contribution to an amplitude samples the form factor through a momentum weight,
\begin{equation}
\langle f\rangle_W=
\frac{\int_0^\infty dp\,p^2W(p)f(p)}
{\int_0^\infty dp\,p^2W(p)},
\label{eq:factorized}
\end{equation}
the remaining reaction factors multiplying $\langle\tilde D_S\rangle_W$. In a full amplitude $W$
carries spin projection and distorted waves and can be complex; the examples below use real,
non-negative weights, characterized by the root-mean-square momentum they sample,
$p_{\rm rms}=\langle p^2\rangle_W^{1/2}$. In the
plane-wave limit of Ref.~\cite{ButtleGoldfarb1964} the weight is concentrated at the transferred
momentum.

The Taylor expansion about $p^2=0$ converges within the nearest singularity in complex momentum.
Once the bound-state singularity at $p^2=-\kappa^2$ is separated, a local one-pion tail places
the next one near
$p=i(\kappa+m_\pi)$, at $\kappa+m_\pi=0.93~\mathrm{fm^{-1}}$; rational fits
to the five form factors estimate the nearest singularity at about 1.0 to $1.1~\mathrm{fm^{-1}}$. This is not the scale
$m_\pi/2$ of the on-shell amplitude, where both nucleons are external and one-pion exchange can
transfer up to $2p$. A finite-order expansion can still approximate a weighted average beyond this
radius if the weighted remainder is small; a small second moment alone does not control the
higher moments or the high-momentum tail.

In coordinate space, with $D_u(r)=-(\kappa^2u_S-u_S'')/m_2$ the radial form factor times $r$,
\begin{equation}
\begin{split}
c_i&=\frac{\int dr\,r^3D_{u,i}}{6\int dr\,r\,D_{u,i}}\equiv\frac{\langle r^2\rangle_{D,i}}{6},\\
\frac{1}{q_i}&=\frac{\int dr\,r\,[\sinh(\kappa_ir)/(\kappa_ir)]\,D_{u,i}}{\int dr\,r\,D_{u,i}},
\end{split}
\label{eq:coordmoments}
\end{equation}
with $q_i$ the ratio to the ANC normalization defined in Sec.~\ref{sec:representation}. The
finite-range coefficient is the signed mean square radius of $D$ divided by six, and the offset
weights $D$ toward large $r$. Evaluated by a sine transform, these moments reproduce the fitted
$c_i$ and $q_i$ to better than $10^{-3}$ relative, for CD Bonn after correcting for the
truncation of its slowly decaying form factor. For the local AV18 interaction $D(r)$ changes sign once,
at 0.65 fm; for the momentum-regulated chiral interactions it keeps oscillating beyond 6 fm, with
a wavelength set by the regulator, so where in $r$ the interactions differ depends on the
regulator. That is one reason to compare them through momentum moments.

The weighted average can be compared with the value fixed by the ANC. With $q_i$ the ratio
defined in Sec.~\ref{sec:representation}, the expansion of $F_i$ gives
\begin{equation}
\delta_{{\rm ANC},i}\equiv1-\frac{\langle\tilde D_{S,i}\rangle_W}{C_iA_{S,i}}
\simeq1-q_i+q_ic_i\,p_{\rm rms}^2,
\label{eq:masteranc}
\end{equation}
with corrections from higher moments of the weight and the curvature of the form factor. The
departure has an offset $1-q_i$, accumulated between $p^2=-\kappa^2$ and zero momentum and
present for every weight, and a slope term. For a common weight the range of the slope term
across the interactions is $p_{\rm rms}^2$ times the range of $q_ic_i$. To compare shapes
independently of the ANC normalization, I use the range of $1-\langle F_i\rangle_W$ across the
interactions and call it the shape range; its leading estimate is $p_{\rm rms}^2\Delta c$, with
$\Delta c$ the range of the finite-range coefficients. For these five interactions the offset and slope add, since $1-q_i$ grows with $c_i$.
The offsets range over 0.457 percent and $\Delta c=0.0843~\mathrm{fm^2}$, 15 to 17 percent of
$c_i$ itself. The coefficients come from a polynomial fit to
$\tilde D_S$ in $p^2$ including $p^4$ and $p^6$, on 200-point meshes.

\section{Where standard reaction treatments sample the form factor}
\label{sec:where}

Table~\ref{tab:weights} defines three weights. The window weight is not the weight of any
reaction: it is a common reference, the same for all interactions, that samples only momenta
below $0.30~\mathrm{fm^{-1}}$, where the low-momentum expansion should hold. The density weight $\rho_i$ samples the deuteron momentum
distribution, as in the Watanabe folding potential $\langle\phi_d|U_{nA}+U_{pA}|\phi_d\rangle$
\cite{Watanabe1958}, the ground-state diagonal potential of a continuum-discretized
coupled-channels (CDCC) calculation before coupling to breakup, which adds a nonlocal
polarization potential that no single weight represents. The third weight carries the factor
$p^2+\kappa^2$ of Eq.~(\ref{eq:identity}), as in the ADWA weight $\bar\phi_1\propto V_{np}\phi_d$
\cite{JohnsonSoper1970,JohnsonTandy1974}, and reproduces
$\langle T_{np}\rangle_V=(\hbar c/m_2)p_{\rm rms}^2$ by construction. The root-mean-square
momenta of the three are 0.23, 0.61 and $1.69~\mathrm{fm^{-1}}$.

\begin{table}[t]
\caption{Weights used to sample the $S$-wave form factor. Here
$\rho_i(p)=|\tilde\phi_{S,i}(p)|^2+|\tilde\phi_{D,i}(p)|^2$ and
$\Theta$ is the unit step function, $p_c=0.30~\mathrm{fm^{-1}}$. The last column gives the root-mean-square momentum
$p_{\rm rms}$ in $\mathrm{fm^{-1}}$, from $\langle p^2\rangle_W$ averaged over the five
interactions for the two interaction-dependent weights. These are structure factors, not full reaction weights.}
\label{tab:weights}
\begin{ruledtabular}
\begin{tabular}{lcc}
Weight & $W_i(p)$ & $p_{\rm rms}$\\
\hline
Window & $\Theta(p_c-p)$ & 0.23\\
Watanabe & $\rho_i(p)$ & 0.61\\
ADWA & $(p^2+\kappa_i^2)\rho_i(p)$ & 1.69\\
\end{tabular}
\end{ruledtabular}
\end{table}

At the window weight the low-momentum expansion holds. Equation~(\ref{eq:masteranc}) predicts
an ANC-normalized range of 0.886 percent across the five interactions against 0.860 percent
from direct integration, and a shape range of 0.455 percent against 0.428 percent; neither
number is fitted to the other, and about half of the ANC-normalized range is already present at
zero momentum. Gaussian weights $W_a(p)=\exp(-p^2/a^2)$ of increasing width depart increasingly
from the leading expression; by $p_{\rm rms}\simeq0.7~\mathrm{fm^{-1}}$ higher-order terms change
the ranges substantially, and the weighted averages of the interactions have begun to cross and
reorder. The Watanabe and ADWA weights move with the interaction,
and their shape ranges, 1.8 and 17 percent, combine a change of sampling with a change of form
factor; their ordering reflects the greater emphasis on short distances in ADWA
\cite{GomezRamosTimofeyuk2018,Timofeyuk2020}. ADWA itself retains the full $V_{np}\phi_d$, so the
failure of this expansion at the ADWA weight limits the expansion, not ADWA. The published tables
make the same contrast: across the four deuteron models of Ref.~\cite{Bailey2017} the
zero-range constant varies by 0.82 percent while $\langle T_{np}\rangle_V$ varies by a factor of
2.3.

The unitary family tests this at the level of ranges: at leading order, the shape range it
produces should equal the model shape range scaled by the ratio of the family and model ranges
of $c$. For the window weight and the two sampled Gaussian weights with
$p_{\rm rms}^2\Delta c<0.01$ it does so to within a factor of 0.98 to 1.12 as $R_h$ varies from
1.2 to 3.0 fm, while the family range of $c$ changes by a factor of 32. Wider weights depart in
either direction once they sample the curvature of the form factors. Agreement at low momentum
follows from matching the slope and is not independent evidence; matching a slope at the origin
places no restriction at large momentum.

\section{Recovering the kernel by consistent transformation}
\label{sec:recovery}

For a folded potential, the change of $V_{np}\phi_d$ under $U$ must be accompanied by changes of
the nucleon-target interactions and of the operator in the matrix element
\cite{PolyzouGlockle1990,Bogner2007,Anderson2010}. With $U_N\equiv U_{nA}+U_{pA}$ the nucleon-target interaction, and since $U$ acts only on
the internal coordinate,
\begin{equation}
\begin{split}
U(T_R+T+V_{np}+U_N)U^\dagger
&=T_R+T+V_\theta+U_N+W_3,\\
W_3&=UU_NU^\dagger-U_N,
\end{split}
\label{eq:embedded}
\end{equation}
and the transformed transition operator is
\begin{equation}
UV_{np}U^\dagger=V_\theta+T-UTU^\dagger.
\label{eq:operatortransform}
\end{equation}
Inserting $V_\theta$ into the original form-factor expression omits the kinetic term of
Eq.~(\ref{eq:operatortransform}); the values in Table~\ref{tab:tnp} contain the same omission.
Because $U-1$ has rank two, $W_3$ has rank at most four at fixed $R$, and its matrix elements
reduce to overlaps with the two rotation functions and their folded images.

I checked the recovery on the normalized $S$-projected folding kernel
$K(R)=\langle\bar\phi_{1,S}|U_N|\phi_{d,S}\rangle/\langle\bar\phi_{1,S}|\phi_{d,S}\rangle$ of the
AV18 deuteron, where $\bar\phi_{1,S}$ is the $S$ projection of the full coupled weight,
$V_{SS}\phi_{d,S}+V_{SD}\phi_{d,D}=-(T+B_d)\phi_{d,S}$, so the tensor coupling to the $D$ wave is
included, with $U_N$ averaged over the direction of the neutron-proton separation,
with a real Woods-Saxon nucleon-target potential of depth $-45.0$ MeV, radius 4.60 fm and
diffuseness 0.65 fm, through the radial integral $J=\int_{0.5\,{\rm fm}}^{12\,{\rm fm}}dR\,K(R)$, at
$R_h=2.0$ fm. Three versions differ only in the left weight and the operator; the right state is
$U\phi_d$ in all three. With the left weight rebuilt as $V_\theta U\phi_d$ and $U_N$ untransformed,
$J$ changes by 0.48 percent at $\theta=0.40$. Adding $W_3$ leaves 0.32 percent, the residual of a
Johnson-Tandy weight rebuilt from $V_\theta$ with the internal kinetic operator held fixed.
Replacing the left weight by the transported $UV_{np}\phi_d$, together with its normalization,
restores the kernel to numerical precision, as unitarity requires. The intermediate residuals
measure the artificial dependence caused by omitting transformations, in the sense of
Ref.~\cite{Duguet2015}. $W_3$ is neither a target-excitation three-body force of a Feshbach
reduction \cite{JohnsonTimofeyuk2014,Dinmore2019,Dinmore2021} nor a fitted three-nucleon force
\cite{Timofeyuk2020}, and the closure places no bound on differences between such Hamiltonians.

\section{High-momentum probes and limits}
\label{sec:scope}

The weighted kinetic energy belongs to the neutron-proton structure factor; beam energy does not set it
by itself, and for AV18 the ADWA weight gives $\langle T_{np}\rangle_V=217.9$ MeV at any incident
energy. In a full amplitude, beam energy, angle, channel coupling and optical potentials change
the weight. A plane-wave reduction of $(p,d)$ pickup makes this explicit: with the zero-degree
kinematics of Hai \textit{et al.}~\cite{Hai2026} the transferred momentum grows from
$0.67~\mathrm{fm^{-1}}$ at 50 MeV to $2.45~\mathrm{fm^{-1}}$ at 800 MeV. At 50 MeV the five
$S$-wave shapes differ by 2 percent; at 600 MeV the $S$-wave
component changes sign among the interactions, the $D$-wave magnitude varies by a factor of four,
and it exceeds the $S$-wave magnitude for four of the five, so the low-momentum expansion fails,
and the largest magnitudes of the two components occur for different interactions.

Measurements of high-momentum nucleon pairs probe this regime \cite{Terashima2018}. A unitary
transformation that leaves every two-body observable unchanged also moves strength into high
momentum, so a spectroscopic amplitude or a fraction of tensor-correlated pairs extracted from
measured cross sections is defined only together with the representation of the states and of
the reaction operator \cite{Duguet2015,More2017,Tropiano2021}.

The calculation covers the deuteron form factor only. Exact three-body calculations find transfer
contributions from neutron-proton partial waves outside the deuteron channel \cite{Deltuva2018},
and target excitation introduces additional uncertainties \cite{Dinmore2021}.

\section{Summary}
\label{sec:summary}

The neutron-proton kinetic energy weighted by the finite-range form factor, one identified
structure integral through which nonlocal ADWA calculations depend on $V_{np}$, equals the mean kinetic
energy plus its variance divided by $\langle T\rangle+B_d$, and the variance term is 79 to
91 percent of the weighted kinetic energy for the five realistic interactions studied. A compactly supported unitary
transformation of the neutron-proton Hamiltonian with the AV18 interaction in the deuteron
channel, preserving the two-body scattering matrix, the binding
energy, and the asymptotic normalizations, raises it from 217.9 to 314.0 MeV, a change equal to
62 percent of the realistic span of 63 to 218 MeV, while the finite-range coefficient moves by one percent of its
realistic range. When the coefficient moves by three percent of its range, the weighted
kinetic energy exceeds every realistic value. A dependence carried by this integral therefore
cannot on its own constrain short-range neutron-proton structure.

The dependence reaches ADWA through two features of the adiabatic projection: the weight
$V_{np}\phi_d$, which moves the nonlocal folding to the separations where $V_{np}$ acts, and
the frozen neutron-proton energy: the internal Hamiltonian is replaced by $-B_d$, while the
internal momentum sampled there enters through the momentum dependence of the nucleon potentials.
With the Perey-Buck factor kept, the range of the Gaussian-averaged overlap with the displaced deuteron is
27 percent of its mean over the five interactions under the adiabatic weight and 1.3 percent under the deuteron density
that sets the continuum-discretized elastic-channel coefficient, and in a Faddeev calculation
energy conservation bounds the nucleon-target energy. Keeping the same factor reduces the
representation dependence of the overlap without removing it: the transformation that moves
$\langle T_{np}\rangle_V$ by 62 percent of its realistic span moves the bounded average
$T_\beta$ by 12 percent of its own realistic span. Expanding the nonlocal kernel to lowest order therefore strongly
amplifies the response of the overlap to a change confined to short distances.

The low-momentum content of the form factor is the finite-range expansion of Buttle and
Goldfarb, here stated for realistic coupled-channel deuterons. At low momentum the shapes of the
realistic form factors differ mainly through their finite-range coefficients, a difference of
$0.51~\mathrm{fm^2}$ in the mean square radius of $D$, and a wider rotation ($R_h=2.0$ fm,
$\theta=0.40$) moves the offset and the coefficient by amounts comparable to their realistic
ranges, so even this low-momentum variation can arise from representation alone.

Consistent transformation induces a three-body term of rank at most four at fixed deuteron-target
separation $\mathbf R$. In the $S$-projected folding example,
transforming states and operators together restores the kernel, while omitting part of the
transformation leaves artificial residuals below one percent.

At the zero-degree momentum sampled in 600-MeV $(p,d)$ pickup, the $S$-wave form factor changes sign
among the realistic interactions and the $D$-wave magnitude varies by a factor of four.
Interpreting a measured reaction in terms of short-distance or tensor structure therefore requires
a representation shared by states and operators and the reaction's own momentum weight. Without
these, a spread of $V_{np}\phi_d$ among interactions is neither a reaction uncertainty nor an
observable constraint on short-distance structure.

\begin{acknowledgments}
This work was supported by the National Natural Science Foundation of China (Grant
Nos.~12475132 and 12535009) and the Fundamental Research Funds for the Central Universities.

Large language models assisted with prose
editing, code review, algebraic consistency checks, and verification
of numerical and bibliographic records. The author is responsible
for the derivations, numerical results, and interpretation.
\end{acknowledgments}

\section*{Data availability}

The analysis scripts that generate every number and figure in this paper, the interaction
tabulations they read, and the mesh definitions needed to reproduce the converged results are
available from the author on reasonable request. The interactions themselves are the published ones
cited in Table~\ref{tab:potentials}.

\bibliography{refs}

\begin{thebibliography}{36}%
\makeatletter
\providecommand \@ifxundefined [1]{%
 \@ifx{#1\undefined}
}%
\providecommand \@ifnum [1]{%
 \ifnum #1\expandafter \@firstoftwo
 \else \expandafter \@secondoftwo
 \fi
}%
\providecommand \@ifx [1]{%
 \ifx #1\expandafter \@firstoftwo
 \else \expandafter \@secondoftwo
 \fi
}%
\providecommand \natexlab [1]{#1}%
\providecommand \enquote  [1]{``#1''}%
\providecommand \bibnamefont  [1]{#1}%
\providecommand \bibfnamefont [1]{#1}%
\providecommand \citenamefont [1]{#1}%
\providecommand \href@noop [0]{\@secondoftwo}%
\providecommand \href [0]{\begingroup \@sanitize@url \@href}%
\providecommand \@href[1]{\@@startlink{#1}\@@href}%
\providecommand \@@href[1]{\endgroup#1\@@endlink}%
\providecommand \@sanitize@url [0]{\catcode `\\12\catcode `\$12\catcode
  `\&12\catcode `\#12\catcode `\^12\catcode `\_12\catcode `\%12\relax}%
\providecommand \@@startlink[1]{}%
\providecommand \@@endlink[0]{}%
\providecommand \url  [0]{\begingroup\@sanitize@url \@url }%
\providecommand \@url [1]{\endgroup\@href {#1}{\urlprefix }}%
\providecommand \urlprefix  [0]{URL }%
\providecommand \Eprint [0]{\href }%
\providecommand \doibase [0]{https://doi.org/}%
\providecommand \selectlanguage [0]{\@gobble}%
\providecommand \bibinfo  [0]{\@secondoftwo}%
\providecommand \bibfield  [0]{\@secondoftwo}%
\providecommand \translation [1]{[#1]}%
\providecommand \BibitemOpen [0]{}%
\providecommand \bibitemStop [0]{}%
\providecommand \bibitemNoStop [0]{.\EOS\space}%
\providecommand \EOS [0]{\spacefactor3000\relax}%
\providecommand \BibitemShut  [1]{\csname bibitem#1\endcsname}%
\let\auto@bib@innerbib\@empty
\bibitem [{\citenamefont {Johnson}\ and\ \citenamefont
  {Soper}(1970)}]{JohnsonSoper1970}%
  \BibitemOpen
  \bibfield  {author} {\bibinfo {author} {\bibfnamefont {R.~C.}\ \bibnamefont
  {Johnson}}\ and\ \bibinfo {author} {\bibfnamefont {P.~J.~R.}\ \bibnamefont
  {Soper}},\ }\href {https://doi.org/10.1103/PhysRevC.1.976} {\bibfield
  {journal} {\bibinfo  {journal} {Phys. Rev. C}\ }\textbf {\bibinfo {volume}
  {1}},\ \bibinfo {pages} {976} (\bibinfo {year} {1970})}\BibitemShut {NoStop}%
\bibitem [{\citenamefont {Johnson}\ and\ \citenamefont
  {Tandy}(1974)}]{JohnsonTandy1974}%
  \BibitemOpen
  \bibfield  {author} {\bibinfo {author} {\bibfnamefont {R.~C.}\ \bibnamefont
  {Johnson}}\ and\ \bibinfo {author} {\bibfnamefont {P.~C.}\ \bibnamefont
  {Tandy}},\ }\href {https://doi.org/10.1016/0375-9474(74)90178-X} {\bibfield
  {journal} {\bibinfo  {journal} {Nucl. Phys. A}\ }\textbf {\bibinfo {volume}
  {235}},\ \bibinfo {pages} {56} (\bibinfo {year} {1974})}\BibitemShut
  {NoStop}%
\bibitem [{\citenamefont {Laid}\ \emph {et~al.}(1993)\citenamefont {Laid},
  \citenamefont {Tostevin},\ and\ \citenamefont {Johnson}}]{Laid1993}%
  \BibitemOpen
  \bibfield  {author} {\bibinfo {author} {\bibfnamefont {A.}~\bibnamefont
  {Laid}}, \bibinfo {author} {\bibfnamefont {J.~A.}\ \bibnamefont {Tostevin}},\
  and\ \bibinfo {author} {\bibfnamefont {R.~C.}\ \bibnamefont {Johnson}},\
  }\href {https://doi.org/10.1103/PhysRevC.48.1307} {\bibfield  {journal}
  {\bibinfo  {journal} {Phys. Rev. C}\ }\textbf {\bibinfo {volume} {48}},\
  \bibinfo {pages} {1307} (\bibinfo {year} {1993})}\BibitemShut {NoStop}%
\bibitem [{\citenamefont {Pang}\ \emph {et~al.}(2013)\citenamefont {Pang},
  \citenamefont {Timofeyuk}, \citenamefont {Johnson},\ and\ \citenamefont
  {Tostevin}}]{Pang2013}%
  \BibitemOpen
  \bibfield  {author} {\bibinfo {author} {\bibfnamefont {D.~Y.}\ \bibnamefont
  {Pang}}, \bibinfo {author} {\bibfnamefont {N.~K.}\ \bibnamefont {Timofeyuk}},
  \bibinfo {author} {\bibfnamefont {R.~C.}\ \bibnamefont {Johnson}},\ and\
  \bibinfo {author} {\bibfnamefont {J.~A.}\ \bibnamefont {Tostevin}},\ }\href
  {https://doi.org/10.1103/PhysRevC.87.064613} {\bibfield  {journal} {\bibinfo
  {journal} {Phys. Rev. C}\ }\textbf {\bibinfo {volume} {87}},\ \bibinfo
  {pages} {064613} (\bibinfo {year} {2013})}\BibitemShut {NoStop}%
\bibitem [{\citenamefont {Tsang}\ \emph {et~al.}(2005)\citenamefont {Tsang},
  \citenamefont {Lee},\ and\ \citenamefont {Lynch}}]{Tsang2005}%
  \BibitemOpen
  \bibfield  {author} {\bibinfo {author} {\bibfnamefont {M.~B.}\ \bibnamefont
  {Tsang}}, \bibinfo {author} {\bibfnamefont {J.}~\bibnamefont {Lee}},\ and\
  \bibinfo {author} {\bibfnamefont {W.~G.}\ \bibnamefont {Lynch}},\ }\href
  {https://doi.org/10.1103/PhysRevLett.95.222501} {\bibfield  {journal}
  {\bibinfo  {journal} {Phys. Rev. Lett.}\ }\textbf {\bibinfo {volume} {95}},\
  \bibinfo {pages} {222501} (\bibinfo {year} {2005})}\BibitemShut {NoStop}%
\bibitem [{\citenamefont {Lee}\ \emph {et~al.}(2006)\citenamefont {Lee},
  \citenamefont {Tostevin}, \citenamefont {Brown}, \citenamefont {Delaunay},
  \citenamefont {Lynch}, \citenamefont {Saelim},\ and\ \citenamefont
  {Tsang}}]{Lee2006}%
  \BibitemOpen
  \bibfield  {author} {\bibinfo {author} {\bibfnamefont {J.}~\bibnamefont
  {Lee}}, \bibinfo {author} {\bibfnamefont {J.~A.}\ \bibnamefont {Tostevin}},
  \bibinfo {author} {\bibfnamefont {B.~A.}\ \bibnamefont {Brown}}, \bibinfo
  {author} {\bibfnamefont {F.}~\bibnamefont {Delaunay}}, \bibinfo {author}
  {\bibfnamefont {W.~G.}\ \bibnamefont {Lynch}}, \bibinfo {author}
  {\bibfnamefont {M.~J.}\ \bibnamefont {Saelim}},\ and\ \bibinfo {author}
  {\bibfnamefont {M.~B.}\ \bibnamefont {Tsang}},\ }\href
  {https://doi.org/10.1103/PhysRevC.73.044608} {\bibfield  {journal} {\bibinfo
  {journal} {Phys. Rev. C}\ }\textbf {\bibinfo {volume} {73}},\ \bibinfo
  {pages} {044608} (\bibinfo {year} {2006})}\BibitemShut {NoStop}%
\bibitem [{\citenamefont {Timofeyuk}\ and\ \citenamefont
  {Johnson}(2020)}]{TimofeyukJohnson2020}%
  \BibitemOpen
  \bibfield  {author} {\bibinfo {author} {\bibfnamefont {N.~K.}\ \bibnamefont
  {Timofeyuk}}\ and\ \bibinfo {author} {\bibfnamefont {R.~C.}\ \bibnamefont
  {Johnson}},\ }\href {https://doi.org/10.1016/j.ppnp.2019.103738} {\bibfield
  {journal} {\bibinfo  {journal} {Prog. Part. Nucl. Phys.}\ }\textbf {\bibinfo
  {volume} {111}},\ \bibinfo {pages} {103738} (\bibinfo {year}
  {2020})}\BibitemShut {NoStop}%
\bibitem [{\citenamefont {Timofeyuk}\ and\ \citenamefont
  {Johnson}(2013{\natexlab{a}})}]{TimofeyukJohnson2013PRL}%
  \BibitemOpen
  \bibfield  {author} {\bibinfo {author} {\bibfnamefont {N.~K.}\ \bibnamefont
  {Timofeyuk}}\ and\ \bibinfo {author} {\bibfnamefont {R.~C.}\ \bibnamefont
  {Johnson}},\ }\href {https://doi.org/10.1103/PhysRevLett.110.112501}
  {\bibfield  {journal} {\bibinfo  {journal} {Phys. Rev. Lett.}\ }\textbf
  {\bibinfo {volume} {110}},\ \bibinfo {pages} {112501} (\bibinfo {year}
  {2013}{\natexlab{a}})}\BibitemShut {NoStop}%
\bibitem [{\citenamefont {Timofeyuk}\ and\ \citenamefont
  {Johnson}(2013{\natexlab{b}})}]{TimofeyukJohnson2013}%
  \BibitemOpen
  \bibfield  {author} {\bibinfo {author} {\bibfnamefont {N.~K.}\ \bibnamefont
  {Timofeyuk}}\ and\ \bibinfo {author} {\bibfnamefont {R.~C.}\ \bibnamefont
  {Johnson}},\ }\href {https://doi.org/10.1103/PhysRevC.87.064610} {\bibfield
  {journal} {\bibinfo  {journal} {Phys. Rev. C}\ }\textbf {\bibinfo {volume}
  {87}},\ \bibinfo {pages} {064610} (\bibinfo {year}
  {2013}{\natexlab{b}})}\BibitemShut {NoStop}%
\bibitem [{\citenamefont {Johnson}\ and\ \citenamefont
  {Timofeyuk}(2014)}]{JohnsonTimofeyuk2014}%
  \BibitemOpen
  \bibfield  {author} {\bibinfo {author} {\bibfnamefont {R.~C.}\ \bibnamefont
  {Johnson}}\ and\ \bibinfo {author} {\bibfnamefont {N.~K.}\ \bibnamefont
  {Timofeyuk}},\ }\href {https://doi.org/10.1103/PhysRevC.89.024605} {\bibfield
   {journal} {\bibinfo  {journal} {Phys. Rev. C}\ }\textbf {\bibinfo {volume}
  {89}},\ \bibinfo {pages} {024605} (\bibinfo {year} {2014})}\BibitemShut
  {NoStop}%
\bibitem [{\citenamefont {Dinmore}\ \emph {et~al.}(2019)\citenamefont
  {Dinmore}, \citenamefont {Timofeyuk}, \citenamefont {Al-Khalili},\ and\
  \citenamefont {Johnson}}]{Dinmore2019}%
  \BibitemOpen
  \bibfield  {author} {\bibinfo {author} {\bibfnamefont {M.~J.}\ \bibnamefont
  {Dinmore}}, \bibinfo {author} {\bibfnamefont {N.~K.}\ \bibnamefont
  {Timofeyuk}}, \bibinfo {author} {\bibfnamefont {J.~S.}\ \bibnamefont
  {Al-Khalili}},\ and\ \bibinfo {author} {\bibfnamefont {R.~C.}\ \bibnamefont
  {Johnson}},\ }\href {https://doi.org/10.1103/PhysRevC.99.064612} {\bibfield
  {journal} {\bibinfo  {journal} {Phys. Rev. C}\ }\textbf {\bibinfo {volume}
  {99}},\ \bibinfo {pages} {064612} (\bibinfo {year} {2019})}\BibitemShut
  {NoStop}%
\bibitem [{\citenamefont {Dinmore}\ \emph {et~al.}(2021)\citenamefont
  {Dinmore}, \citenamefont {Timofeyuk},\ and\ \citenamefont
  {Al-Khalili}}]{Dinmore2021}%
  \BibitemOpen
  \bibfield  {author} {\bibinfo {author} {\bibfnamefont {M.~J.}\ \bibnamefont
  {Dinmore}}, \bibinfo {author} {\bibfnamefont {N.~K.}\ \bibnamefont
  {Timofeyuk}},\ and\ \bibinfo {author} {\bibfnamefont {J.~S.}\ \bibnamefont
  {Al-Khalili}},\ }\href {https://doi.org/10.1103/PhysRevC.104.034614}
  {\bibfield  {journal} {\bibinfo  {journal} {Phys. Rev. C}\ }\textbf {\bibinfo
  {volume} {104}},\ \bibinfo {pages} {034614} (\bibinfo {year}
  {2021})}\BibitemShut {NoStop}%
\bibitem [{\citenamefont {Timofeyuk}\ \emph {et~al.}(2020)\citenamefont
  {Timofeyuk}, \citenamefont {Dinmore},\ and\ \citenamefont
  {Al-Khalili}}]{Timofeyuk2020}%
  \BibitemOpen
  \bibfield  {author} {\bibinfo {author} {\bibfnamefont {N.~K.}\ \bibnamefont
  {Timofeyuk}}, \bibinfo {author} {\bibfnamefont {M.~J.}\ \bibnamefont
  {Dinmore}},\ and\ \bibinfo {author} {\bibfnamefont {J.~S.}\ \bibnamefont
  {Al-Khalili}},\ }\href {https://doi.org/10.1103/PhysRevC.102.064616}
  {\bibfield  {journal} {\bibinfo  {journal} {Phys. Rev. C}\ }\textbf {\bibinfo
  {volume} {102}},\ \bibinfo {pages} {064616} (\bibinfo {year}
  {2020})}\BibitemShut {NoStop}%
\bibitem [{\citenamefont {Bailey}\ \emph {et~al.}(2016)\citenamefont {Bailey},
  \citenamefont {Timofeyuk},\ and\ \citenamefont {Tostevin}}]{Bailey2016}%
  \BibitemOpen
  \bibfield  {author} {\bibinfo {author} {\bibfnamefont {G.~W.}\ \bibnamefont
  {Bailey}}, \bibinfo {author} {\bibfnamefont {N.~K.}\ \bibnamefont
  {Timofeyuk}},\ and\ \bibinfo {author} {\bibfnamefont {J.~A.}\ \bibnamefont
  {Tostevin}},\ }\href {https://doi.org/10.1103/PhysRevLett.117.162502}
  {\bibfield  {journal} {\bibinfo  {journal} {Phys. Rev. Lett.}\ }\textbf
  {\bibinfo {volume} {117}},\ \bibinfo {pages} {162502} (\bibinfo {year}
  {2016})}\BibitemShut {NoStop}%
\bibitem [{\citenamefont {Bailey}\ \emph
  {et~al.}(2017{\natexlab{a}})\citenamefont {Bailey}, \citenamefont
  {Timofeyuk},\ and\ \citenamefont {Tostevin}}]{Bailey2017erratum}%
  \BibitemOpen
  \bibfield  {author} {\bibinfo {author} {\bibfnamefont {G.~W.}\ \bibnamefont
  {Bailey}}, \bibinfo {author} {\bibfnamefont {N.~K.}\ \bibnamefont
  {Timofeyuk}},\ and\ \bibinfo {author} {\bibfnamefont {J.~A.}\ \bibnamefont
  {Tostevin}},\ }\href {https://doi.org/10.1103/PhysRevLett.119.159901}
  {\bibfield  {journal} {\bibinfo  {journal} {Phys. Rev. Lett.}\ }\textbf
  {\bibinfo {volume} {119}},\ \bibinfo {pages} {159901} (\bibinfo {year}
  {2017}{\natexlab{a}})}\BibitemShut {NoStop}%
\bibitem [{\citenamefont {Bailey}\ \emph
  {et~al.}(2017{\natexlab{b}})\citenamefont {Bailey}, \citenamefont
  {Timofeyuk},\ and\ \citenamefont {Tostevin}}]{Bailey2017}%
  \BibitemOpen
  \bibfield  {author} {\bibinfo {author} {\bibfnamefont {G.~W.}\ \bibnamefont
  {Bailey}}, \bibinfo {author} {\bibfnamefont {N.~K.}\ \bibnamefont
  {Timofeyuk}},\ and\ \bibinfo {author} {\bibfnamefont {J.~A.}\ \bibnamefont
  {Tostevin}},\ }\href {https://doi.org/10.1103/PhysRevC.95.024603} {\bibfield
  {journal} {\bibinfo  {journal} {Phys. Rev. C}\ }\textbf {\bibinfo {volume}
  {95}},\ \bibinfo {pages} {024603} (\bibinfo {year}
  {2017}{\natexlab{b}})}\BibitemShut {NoStop}%
\bibitem [{\citenamefont {Deltuva}(2018)}]{Deltuva2018}%
  \BibitemOpen
  \bibfield  {author} {\bibinfo {author} {\bibfnamefont {A.}~\bibnamefont
  {Deltuva}},\ }\href {https://doi.org/10.1103/PhysRevC.98.021603} {\bibfield
  {journal} {\bibinfo  {journal} {Phys. Rev. C}\ }\textbf {\bibinfo {volume}
  {98}},\ \bibinfo {pages} {021603} (\bibinfo {year} {2018})}\BibitemShut
  {NoStop}%
\bibitem [{\citenamefont {G{\'o}mez-Ramos}\ and\ \citenamefont
  {Timofeyuk}(2018)}]{GomezRamosTimofeyuk2018}%
  \BibitemOpen
  \bibfield  {author} {\bibinfo {author} {\bibfnamefont {M.}~\bibnamefont
  {G{\'o}mez-Ramos}}\ and\ \bibinfo {author} {\bibfnamefont {N.~K.}\
  \bibnamefont {Timofeyuk}},\ }\href
  {https://doi.org/10.1103/PhysRevC.98.011601} {\bibfield  {journal} {\bibinfo
  {journal} {Phys. Rev. C}\ }\textbf {\bibinfo {volume} {98}},\ \bibinfo
  {pages} {011601} (\bibinfo {year} {2018})}\BibitemShut {NoStop}%
\bibitem [{\citenamefont {Wiringa}\ \emph {et~al.}(1995)\citenamefont
  {Wiringa}, \citenamefont {Stoks},\ and\ \citenamefont
  {Schiavilla}}]{Wiringa1995}%
  \BibitemOpen
  \bibfield  {author} {\bibinfo {author} {\bibfnamefont {R.~B.}\ \bibnamefont
  {Wiringa}}, \bibinfo {author} {\bibfnamefont {V.~G.~J.}\ \bibnamefont
  {Stoks}},\ and\ \bibinfo {author} {\bibfnamefont {R.}~\bibnamefont
  {Schiavilla}},\ }\href {https://doi.org/10.1103/PhysRevC.51.38} {\bibfield
  {journal} {\bibinfo  {journal} {Phys. Rev. C}\ }\textbf {\bibinfo {volume}
  {51}},\ \bibinfo {pages} {38} (\bibinfo {year} {1995})}\BibitemShut {NoStop}%
\bibitem [{\citenamefont {Buttle}\ and\ \citenamefont
  {Goldfarb}(1964)}]{ButtleGoldfarb1964}%
  \BibitemOpen
  \bibfield  {author} {\bibinfo {author} {\bibfnamefont {P.~J.~A.}\
  \bibnamefont {Buttle}}\ and\ \bibinfo {author} {\bibfnamefont {L.~J.~B.}\
  \bibnamefont {Goldfarb}},\ }\href
  {https://doi.org/10.1088/0370-1328/83/5/302} {\bibfield  {journal} {\bibinfo
  {journal} {Proc. Phys. Soc.}\ }\textbf {\bibinfo {volume} {83}},\ \bibinfo
  {pages} {701} (\bibinfo {year} {1964})}\BibitemShut {NoStop}%
\bibitem [{\citenamefont {Polyzou}\ and\ \citenamefont
  {Gl{\"o}ckle}(1990)}]{PolyzouGlockle1990}%
  \BibitemOpen
  \bibfield  {author} {\bibinfo {author} {\bibfnamefont {W.~N.}\ \bibnamefont
  {Polyzou}}\ and\ \bibinfo {author} {\bibfnamefont {W.}~\bibnamefont
  {Gl{\"o}ckle}},\ }\href {https://doi.org/10.1007/BF01091701} {\bibfield
  {journal} {\bibinfo  {journal} {Few-Body Syst.}\ }\textbf {\bibinfo {volume}
  {9}},\ \bibinfo {pages} {97} (\bibinfo {year} {1990})}\BibitemShut {NoStop}%
\bibitem [{\citenamefont {Duguet}\ \emph {et~al.}(2015)\citenamefont {Duguet},
  \citenamefont {Hergert}, \citenamefont {Holt},\ and\ \citenamefont
  {Som{\`a}}}]{Duguet2015}%
  \BibitemOpen
  \bibfield  {author} {\bibinfo {author} {\bibfnamefont {T.}~\bibnamefont
  {Duguet}}, \bibinfo {author} {\bibfnamefont {H.}~\bibnamefont {Hergert}},
  \bibinfo {author} {\bibfnamefont {J.~D.}\ \bibnamefont {Holt}},\ and\
  \bibinfo {author} {\bibfnamefont {V.}~\bibnamefont {Som{\`a}}},\ }\href
  {https://doi.org/10.1103/PhysRevC.92.034313} {\bibfield  {journal} {\bibinfo
  {journal} {Phys. Rev. C}\ }\textbf {\bibinfo {volume} {92}},\ \bibinfo
  {pages} {034313} (\bibinfo {year} {2015})}\BibitemShut {NoStop}%
\bibitem [{\citenamefont {More}\ \emph {et~al.}(2017)\citenamefont {More},
  \citenamefont {Bogner},\ and\ \citenamefont {Furnstahl}}]{More2017}%
  \BibitemOpen
  \bibfield  {author} {\bibinfo {author} {\bibfnamefont {S.~N.}\ \bibnamefont
  {More}}, \bibinfo {author} {\bibfnamefont {S.~K.}\ \bibnamefont {Bogner}},\
  and\ \bibinfo {author} {\bibfnamefont {R.~J.}\ \bibnamefont {Furnstahl}},\
  }\href {https://doi.org/10.1103/PhysRevC.96.054004} {\bibfield  {journal}
  {\bibinfo  {journal} {Phys. Rev. C}\ }\textbf {\bibinfo {volume} {96}},\
  \bibinfo {pages} {054004} (\bibinfo {year} {2017})}\BibitemShut {NoStop}%
\bibitem [{\citenamefont {Epelbaum}\ \emph {et~al.}(2005)\citenamefont
  {Epelbaum}, \citenamefont {Gl{\"o}ckle},\ and\ \citenamefont
  {Mei{\ss}ner}}]{EpelbaumGlockleMeissner2005}%
  \BibitemOpen
  \bibfield  {author} {\bibinfo {author} {\bibfnamefont {E.}~\bibnamefont
  {Epelbaum}}, \bibinfo {author} {\bibfnamefont {W.}~\bibnamefont
  {Gl{\"o}ckle}},\ and\ \bibinfo {author} {\bibfnamefont {U.-G.}\ \bibnamefont
  {Mei{\ss}ner}},\ }\href {https://doi.org/10.1016/j.nuclphysa.2004.09.107}
  {\bibfield  {journal} {\bibinfo  {journal} {Nucl. Phys. A}\ }\textbf
  {\bibinfo {volume} {747}},\ \bibinfo {pages} {362} (\bibinfo {year}
  {2005})}\BibitemShut {NoStop}%
\bibitem [{\citenamefont {Machleidt}(2001)}]{Machleidt2001}%
  \BibitemOpen
  \bibfield  {author} {\bibinfo {author} {\bibfnamefont {R.}~\bibnamefont
  {Machleidt}},\ }\href {https://doi.org/10.1103/PhysRevC.63.024001} {\bibfield
   {journal} {\bibinfo  {journal} {Phys. Rev. C}\ }\textbf {\bibinfo {volume}
  {63}},\ \bibinfo {pages} {024001} (\bibinfo {year} {2001})}\BibitemShut
  {NoStop}%
\bibitem [{\citenamefont {Entem}\ and\ \citenamefont
  {Machleidt}(2003)}]{EntemMachleidt2003}%
  \BibitemOpen
  \bibfield  {author} {\bibinfo {author} {\bibfnamefont {D.~R.}\ \bibnamefont
  {Entem}}\ and\ \bibinfo {author} {\bibfnamefont {R.}~\bibnamefont
  {Machleidt}},\ }\href {https://doi.org/10.1103/PhysRevC.68.041001} {\bibfield
   {journal} {\bibinfo  {journal} {Phys. Rev. C}\ }\textbf {\bibinfo {volume}
  {68}},\ \bibinfo {pages} {041001} (\bibinfo {year} {2003})}\BibitemShut
  {NoStop}%
\bibitem [{\citenamefont {Capel}\ \emph {et~al.}(2018)\citenamefont {Capel},
  \citenamefont {Phillips},\ and\ \citenamefont {Hammer}}]{Capel2018}%
  \BibitemOpen
  \bibfield  {author} {\bibinfo {author} {\bibfnamefont {P.}~\bibnamefont
  {Capel}}, \bibinfo {author} {\bibfnamefont {D.~R.}\ \bibnamefont
  {Phillips}},\ and\ \bibinfo {author} {\bibfnamefont {H.-W.}\ \bibnamefont
  {Hammer}},\ }\href {https://doi.org/10.1103/PhysRevC.98.034610} {\bibfield
  {journal} {\bibinfo  {journal} {Phys. Rev. C}\ }\textbf {\bibinfo {volume}
  {98}},\ \bibinfo {pages} {034610} (\bibinfo {year} {2018})}\BibitemShut
  {NoStop}%
\bibitem [{\citenamefont {Delaunay}\ \emph {et~al.}(2005)\citenamefont
  {Delaunay}, \citenamefont {Nunes}, \citenamefont {Lynch},\ and\ \citenamefont
  {Tsang}}]{Delaunay2005}%
  \BibitemOpen
  \bibfield  {author} {\bibinfo {author} {\bibfnamefont {F.}~\bibnamefont
  {Delaunay}}, \bibinfo {author} {\bibfnamefont {F.~M.}\ \bibnamefont {Nunes}},
  \bibinfo {author} {\bibfnamefont {W.~G.}\ \bibnamefont {Lynch}},\ and\
  \bibinfo {author} {\bibfnamefont {M.~B.}\ \bibnamefont {Tsang}},\ }\href
  {https://doi.org/10.1103/PhysRevC.72.014610} {\bibfield  {journal} {\bibinfo
  {journal} {Phys. Rev. C}\ }\textbf {\bibinfo {volume} {72}},\ \bibinfo
  {pages} {014610} (\bibinfo {year} {2005})},\ \Eprint
  {https://arxiv.org/abs/nucl-th/0504045} {nucl-th/0504045} \BibitemShut
  {NoStop}%
\bibitem [{\citenamefont {Nunes}\ and\ \citenamefont
  {Deltuva}(2011)}]{NunesDeltuva2011}%
  \BibitemOpen
  \bibfield  {author} {\bibinfo {author} {\bibfnamefont {F.~M.}\ \bibnamefont
  {Nunes}}\ and\ \bibinfo {author} {\bibfnamefont {A.}~\bibnamefont
  {Deltuva}},\ }\href {https://doi.org/10.1103/PhysRevC.84.034607} {\bibfield
  {journal} {\bibinfo  {journal} {Phys. Rev. C}\ }\textbf {\bibinfo {volume}
  {84}},\ \bibinfo {pages} {034607} (\bibinfo {year} {2011})},\ \Eprint
  {https://arxiv.org/abs/1108.2519} {1108.2519} \BibitemShut {NoStop}%
\bibitem [{\citenamefont {Shubhchintak}\ and\ \citenamefont
  {Descouvemont}(2019)}]{Shubhchintak2019}%
  \BibitemOpen
  \bibfield  {author} {\bibinfo {author} {\bibnamefont {Shubhchintak}}\ and\
  \bibinfo {author} {\bibfnamefont {P.}~\bibnamefont {Descouvemont}},\ }\href
  {https://doi.org/10.1103/PhysRevC.100.034611} {\bibfield  {journal} {\bibinfo
   {journal} {Phys. Rev. C}\ }\textbf {\bibinfo {volume} {100}},\ \bibinfo
  {pages} {034611} (\bibinfo {year} {2019})},\ \Eprint
  {https://arxiv.org/abs/1909.02311} {1909.02311} \BibitemShut {NoStop}%
\bibitem [{\citenamefont {Watanabe}(1958)}]{Watanabe1958}%
  \BibitemOpen
  \bibfield  {author} {\bibinfo {author} {\bibfnamefont {S.}~\bibnamefont
  {Watanabe}},\ }\href {https://doi.org/10.1016/0029-5582(58)90180-9}
  {\bibfield  {journal} {\bibinfo  {journal} {Nucl. Phys.}\ }\textbf {\bibinfo
  {volume} {8}},\ \bibinfo {pages} {484} (\bibinfo {year} {1958})}\BibitemShut
  {NoStop}%
\bibitem [{\citenamefont {Bogner}\ \emph {et~al.}(2007)\citenamefont {Bogner},
  \citenamefont {Furnstahl},\ and\ \citenamefont {Perry}}]{Bogner2007}%
  \BibitemOpen
  \bibfield  {author} {\bibinfo {author} {\bibfnamefont {S.~K.}\ \bibnamefont
  {Bogner}}, \bibinfo {author} {\bibfnamefont {R.~J.}\ \bibnamefont
  {Furnstahl}},\ and\ \bibinfo {author} {\bibfnamefont {R.~J.}\ \bibnamefont
  {Perry}},\ }\href {https://doi.org/10.1103/PhysRevC.75.061001} {\bibfield
  {journal} {\bibinfo  {journal} {Phys. Rev. C}\ }\textbf {\bibinfo {volume}
  {75}},\ \bibinfo {pages} {061001} (\bibinfo {year} {2007})}\BibitemShut
  {NoStop}%
\bibitem [{\citenamefont {Anderson}\ \emph {et~al.}(2010)\citenamefont
  {Anderson}, \citenamefont {Bogner}, \citenamefont {Furnstahl},\ and\
  \citenamefont {Perry}}]{Anderson2010}%
  \BibitemOpen
  \bibfield  {author} {\bibinfo {author} {\bibfnamefont {E.~R.}\ \bibnamefont
  {Anderson}}, \bibinfo {author} {\bibfnamefont {S.~K.}\ \bibnamefont
  {Bogner}}, \bibinfo {author} {\bibfnamefont {R.~J.}\ \bibnamefont
  {Furnstahl}},\ and\ \bibinfo {author} {\bibfnamefont {R.~J.}\ \bibnamefont
  {Perry}},\ }\href {https://doi.org/10.1103/PhysRevC.82.054001} {\bibfield
  {journal} {\bibinfo  {journal} {Phys. Rev. C}\ }\textbf {\bibinfo {volume}
  {82}},\ \bibinfo {pages} {054001} (\bibinfo {year} {2010})}\BibitemShut
  {NoStop}%
\bibitem [{\citenamefont {Hai}\ \emph {et~al.}(2026)\citenamefont {Hai},
  \citenamefont {Pang}, \citenamefont {Tanihata}, \citenamefont {Ong},
  \citenamefont {Terashima}, \citenamefont {Wang}, \citenamefont {Xu},
  \citenamefont {Chen}, \citenamefont {Chen},\ and\ \citenamefont
  {Yan}}]{Hai2026}%
  \BibitemOpen
  \bibfield  {author} {\bibinfo {author} {\bibfnamefont {W.~L.}\ \bibnamefont
  {Hai}}, \bibinfo {author} {\bibfnamefont {D.~Y.}\ \bibnamefont {Pang}},
  \bibinfo {author} {\bibfnamefont {I.}~\bibnamefont {Tanihata}}, \bibinfo
  {author} {\bibfnamefont {H.~J.}\ \bibnamefont {Ong}}, \bibinfo {author}
  {\bibfnamefont {S.}~\bibnamefont {Terashima}}, \bibinfo {author}
  {\bibfnamefont {X.}~\bibnamefont {Wang}}, \bibinfo {author} {\bibfnamefont
  {Y.~P.}\ \bibnamefont {Xu}}, \bibinfo {author} {\bibfnamefont {W.~D.}\
  \bibnamefont {Chen}}, \bibinfo {author} {\bibfnamefont {R.~Y.}\ \bibnamefont
  {Chen}},\ and\ \bibinfo {author} {\bibfnamefont {J.~J.}\ \bibnamefont
  {Yan}},\ }\href@noop {} {\bibinfo {title} {Implications of relativistic
  corrections on high-momentum nucleon-transfer reactions}} (\bibinfo {year}
  {2026}),\ \Eprint {https://arxiv.org/abs/2608.28417} {arXiv:2608.28417
  [nucl-th]} \BibitemShut {NoStop}%
\bibitem [{\citenamefont {Terashima}\ \emph {et~al.}(2018)\citenamefont
  {Terashima}, \citenamefont {Yu}, \citenamefont {Ong}, \citenamefont
  {Tanihata}, \citenamefont {Adachi}, \citenamefont {Aoi}, \citenamefont
  {Chan}, \citenamefont {Fujioka}, \citenamefont {Fukuda}, \citenamefont
  {Geissel}, \citenamefont {Gey}, \citenamefont {Golak}, \citenamefont
  {Haettner}, \citenamefont {Iwamoto}, \citenamefont {Kawabata}, \citenamefont
  {Kamada}, \citenamefont {Le}, \citenamefont {Sakaguchi}, \citenamefont
  {Sakaue}, \citenamefont {Scheidenberger}, \citenamefont {Skibi{\'n}ski},
  \citenamefont {Sun}, \citenamefont {Tamii}, \citenamefont {Tang},
  \citenamefont {Tran}, \citenamefont {Topolnicki}, \citenamefont {Wang},
  \citenamefont {Watanabe}, \citenamefont {Weick}, \citenamefont {Wita{\l}a},
  \citenamefont {Zhang},\ and\ \citenamefont {Zhu}}]{Terashima2018}%
  \BibitemOpen
  \bibfield  {author} {\bibinfo {author} {\bibfnamefont {S.}~\bibnamefont
  {Terashima}}, \bibinfo {author} {\bibfnamefont {L.}~\bibnamefont {Yu}},
  \bibinfo {author} {\bibfnamefont {H.~J.}\ \bibnamefont {Ong}}, \bibinfo
  {author} {\bibfnamefont {I.}~\bibnamefont {Tanihata}}, \bibinfo {author}
  {\bibfnamefont {S.}~\bibnamefont {Adachi}}, \bibinfo {author} {\bibfnamefont
  {N.}~\bibnamefont {Aoi}}, \bibinfo {author} {\bibfnamefont {P.~Y.}\
  \bibnamefont {Chan}}, \bibinfo {author} {\bibfnamefont {H.}~\bibnamefont
  {Fujioka}}, \bibinfo {author} {\bibfnamefont {M.}~\bibnamefont {Fukuda}},
  \bibinfo {author} {\bibfnamefont {H.}~\bibnamefont {Geissel}}, \bibinfo
  {author} {\bibfnamefont {G.}~\bibnamefont {Gey}}, \bibinfo {author}
  {\bibfnamefont {J.}~\bibnamefont {Golak}}, \bibinfo {author} {\bibfnamefont
  {E.}~\bibnamefont {Haettner}}, \bibinfo {author} {\bibfnamefont
  {C.}~\bibnamefont {Iwamoto}}, \bibinfo {author} {\bibfnamefont
  {T.}~\bibnamefont {Kawabata}}, \bibinfo {author} {\bibfnamefont
  {H.}~\bibnamefont {Kamada}}, \bibinfo {author} {\bibfnamefont {X.~Y.}\
  \bibnamefont {Le}}, \bibinfo {author} {\bibfnamefont {H.}~\bibnamefont
  {Sakaguchi}}, \bibinfo {author} {\bibfnamefont {A.}~\bibnamefont {Sakaue}},
  \bibinfo {author} {\bibfnamefont {C.}~\bibnamefont {Scheidenberger}},
  \bibinfo {author} {\bibfnamefont {R.}~\bibnamefont {Skibi{\'n}ski}}, \bibinfo
  {author} {\bibfnamefont {B.~H.}\ \bibnamefont {Sun}}, \bibinfo {author}
  {\bibfnamefont {A.}~\bibnamefont {Tamii}}, \bibinfo {author} {\bibfnamefont
  {T.~L.}\ \bibnamefont {Tang}}, \bibinfo {author} {\bibfnamefont {D.~T.}\
  \bibnamefont {Tran}}, \bibinfo {author} {\bibfnamefont {K.}~\bibnamefont
  {Topolnicki}}, \bibinfo {author} {\bibfnamefont {T.~F.}\ \bibnamefont
  {Wang}}, \bibinfo {author} {\bibfnamefont {Y.~N.}\ \bibnamefont {Watanabe}},
  \bibinfo {author} {\bibfnamefont {H.}~\bibnamefont {Weick}}, \bibinfo
  {author} {\bibfnamefont {H.}~\bibnamefont {Wita{\l}a}}, \bibinfo {author}
  {\bibfnamefont {G.~X.}\ \bibnamefont {Zhang}},\ and\ \bibinfo {author}
  {\bibfnamefont {L.~H.}\ \bibnamefont {Zhu}},\ }\href
  {https://doi.org/10.1103/PhysRevLett.121.242501} {\bibfield  {journal}
  {\bibinfo  {journal} {Phys. Rev. Lett.}\ }\textbf {\bibinfo {volume} {121}},\
  \bibinfo {pages} {242501} (\bibinfo {year} {2018})}\BibitemShut {NoStop}%
\bibitem [{\citenamefont {Tropiano}\ \emph {et~al.}(2021)\citenamefont
  {Tropiano}, \citenamefont {Bogner},\ and\ \citenamefont
  {Furnstahl}}]{Tropiano2021}%
  \BibitemOpen
  \bibfield  {author} {\bibinfo {author} {\bibfnamefont {A.~J.}\ \bibnamefont
  {Tropiano}}, \bibinfo {author} {\bibfnamefont {S.~K.}\ \bibnamefont
  {Bogner}},\ and\ \bibinfo {author} {\bibfnamefont {R.~J.}\ \bibnamefont
  {Furnstahl}},\ }\href {https://doi.org/10.1103/PhysRevC.104.034311}
  {\bibfield  {journal} {\bibinfo  {journal} {Phys. Rev. C}\ }\textbf {\bibinfo
  {volume} {104}},\ \bibinfo {pages} {034311} (\bibinfo {year} {2021})},\
  \Eprint {https://arxiv.org/abs/2105.13936} {arXiv:2105.13936 [nucl-th]}
  \BibitemShut {NoStop}%
\end{thebibliography}%

\end{document}